\documentclass[english,reprint,aps]{revtex4-1}
\usepackage[T1]{fontenc}
\usepackage[latin9]{inputenc}
\usepackage{babel}
\usepackage{amsmath}
\usepackage{amssymb}
\usepackage{graphicx}
\usepackage{eqparbox}
\usepackage[unicode=true,pdfusetitle,
 bookmarks=true,bookmarksnumbered=true,bookmarksopen=true,bookmarksopenlevel=1,
 breaklinks=true,pdfborder={0 0 1},backref=false,colorlinks=false]
 {hyperref}

\makeatletter

\providecommand{\tabularnewline}{\\}

\makeatother

\begin{document}

\preprint{This line only printed with preprint option}

\title{Minimal QCDF Model for $B\to\pi K$ Puzzle}

\author{Tsung-Wen Yeh}

\affiliation{Department of Science Education and Application,National Taichung
University of Education, Taiwan, R.O.C.}
\email{twyeh@mail.ntcu.edu.tw}

\date{\today}
\begin{abstract}
In this work, we propose a model by combining the parametrization method and the QCD factorization to study the $B\to\pi K$ puzzle.
The parametrization for the $B\to\pi K$ amplitudes introduces twelve
parameters, the weak angle $\gamma$ and the other eleven hadronic parameters.
These hadronic parameters are assumed to be perturbative and can be calculated by the QCD factorization (QCDF).
The calculational accuracy of the QCDF is improved by including the twist-3 three parton tree and one loop corrections.
Three additional nonperturbative strong phases from the gluonic penguin, color suppressed, and color favored tree diagrams
are required to account for the direct CP asymmetries.
The weak angle, three nonperturbative strong phases, and four scale variables are assumed as fitting parameters.
Four scale variables represent the factorization scales for the decays, each decay mode with one scale.
These eight parameters are determined by a least squares fit to the eight measurements for $B\to\pi K$ decays,
four branching ratios and four direct CP asymmetries.
The fit shows that the hadronic parameters need to be process-dependent.
A large negative relative phase associated with the internal up quark gluonic penguin diagram is observed.
The weak angle $\gamma$ is found to be $72.1\pm5.8$ degree,
being consistent with the world averaged value, $72.1\pm5.8$ degree.
By a least squares fit to the mixing induced asymmetry $S(\pi^{0}K_{S})=0.58\pm0.06$,
the weak angle $\beta$ is determined to be $22.9\pm2.1$ degree,
which is in a well agreement with the world averaged value, $22.5\pm4.4\pm0.6$ degree.
The model predicts the ratio $C^{\prime}/T^{\prime}$
(color suppressed/color favored) to be $0.59\pm0.08$.
The above evidences indicate that the model can solve the $B\to\pi K$ puzzle consistently.
The model is used to examine the quadrangle relation of the isospin symmetry assumed for the $B\to\pi K$ system.
The ratio of the two sides of the quadrangle relation is calculated to be $(0.87\pm0.15)\exp(-i22^{\circ})$,
which signals the isospin symmetry breaking at $6\sigma$.
The process-dependent hadronic parameters break the isospin symmetry ``dynamically''.
The application of our method to other decay processes is straightforward.
\begin{description}
\item [{PACS~numbers}] 12.15.Ff,12.38.Bx,12.39.St,13.25.Hw, 14.40.Nd.{\small \par}
\item [{Keywords}] perturbative QCD, factorization, isospin, B physics,
CP violation, CKM mechanism.{\small \par}
\end{description}
\end{abstract}
\maketitle

\section{\label{sec:Intro}Introduction}

In order to establish the CKM mechanism of the Standard Model (SM),
the studies of hadronic $B$ decays have been devoted to construct
the unitarity triangle $V_{ub}^{*}V_{ud}+V_{cb}^{*}V_{cd}+V_{tb}^{*}V_{td}=0$,
for which a systematic procedure has been developed \citep{2003hep.ph....4132B,2005hep.ph....5175B}.
Experimentally, many rare processes have been measured by BABAR, Belle and LHCb\citep{2014EPJC...74.3026B}.
The benchmarks are the mixing induced CP asymmetry $S(B\to J/\psi K)=0.691\pm0.017$
and the direct CP asymmetry $A_{CP}(B^{0}\to\pi^{-}K^{+})=-0.082\pm0.006$ \citep{Patrignani:2016xqp}.
However, there appeared many difficulties in theories.
One of these is the $B\to\pi K$ puzzle:
\begin{enumerate}
\item The theoretical predictions for the branching ratios are smaller than
the experimental data \citep{BENEKE2001245,Keum:2000wi,2011PhRvD..83c4023L,2014ChPhC..38c3101B}.
\item The theoretical predictions for the CP asymmetries have a much different
pattern from that of the experimental measurements \citep{2011PhRvD..83c4023L,BENEKE2001245,Li:2005kt,Keum:2000wi,2014ChPhC..38c3101B,Khalil:2009zf}.
\item It is difficult to reach consistent explanations for different
experimental measurements \citep{Buras:2003dj,Buras:2003yc,Buras:2004ub}.
\end{enumerate}

At present, it seems still unable to solve the puzzle by model-independent
methods \citep{1126-6708-1999-02-014}, such as the flavor symmetry
approach \citep{Buras:2003dj,Buras:2003yc,Buras:2004ub,Chiang:2004nm,Fleischer:2008wb},
or the global-fit approach \citep{1126-6708-1999-02-014,Chiang:2004nm,Baek:2007yy,Baek:2009pa,Beaudry2018}.
The other possible method could be to employ the perturbative QCD
(pQCD) theories \citep{BENEKE2001245,Keum:2000wi,2011PhRvD..83c4023L,2014ChPhC..38c3101B,2011PhRvD..83c4023L,BENEKE2001245,Li:2005kt,Keum:2000wi,2014ChPhC..38c3101B,Khalil:2009zf}.
In this work, QCD factorization (QCDF)\citep{1999PhRvL..83.1914B,2003NuPhB.675..333B,BENEKE2001245}
is employed to provide a calculational framework.
The concern will focus on how to disentangle the weak phases from the strong phases.
Specifically, a model based on a general parametrization for
the $B\to\pi K$ decay amplitudes will be constructed.
There will introduce eleven hadronic parameters to be calculated by the QCDF.
Three additional nonperturbative strong phases,
which are closely related to the direct CP asymmetries,
are required to explain the experimental data.

In addition, the model assumes that the parameters could be process-dependent.
The reason is easy to understand from the point of view of the pQCD approach.
Any parameter is expressed as a convolution integral of the short and long distance parts.
Although the long distance part is process-independent, however,
the process-dependence of the short distance part makes the hadronic parameter to be process-dependent.
This means that the value of any parameter for different processes would be different.
We will show that this point is important to find a solution to the puzzle.

In past years, the QCDF calculations have considered corrections up to twist-3 two parton and NNLO in $\alpha_{s}$
\citep{BENEKE2001245,2009NuPhB.822..172B,2010NuPhB.832..109B,2015PhLB..750..348B,1126-6708-2002-02-028}.
These calculations show no significant enhancements in the predictions
for the branching ratios and direct CP asymmetries \citep{BENEKE2001245}.
Since the $B\to\pi K$ decays are penguin dominated,
there are twist-3 chirally enhanced corrections from the effective operator $Q_{6,8}$,
for which there also exist three parton contributions.
As indicated in Ref.\citep{Yeh:2008xc}, the tree level three parton
terms could significantly improve the predictions for branching ratios.
In order to improve the calculational accuracy,
we will calculate the three parton contributions up-to one loop radiative corrections.

The organization is as follows. In Sec.\ref{sec:Puzzle-and-method},
a general parametrization method for the amplitudes \citep{2007IJMPA..22.2057I,2000NuPhB.569....3B}
will be used to construct the model for the $B\to\pi K$ decays.
The model contains twelve parameters: the weak angle $\gamma$ and the
other eleven parameters. An improved factorization formula
up-to $O(\alpha_{s}^{2}/m_{b})$ will be constructed by including the
three parton corrections. The formula will be used to calculate the
parameters in Sec\ref{sec:Calculations}.
In Sec\ref{sec:Analysis}, a fitting strategy will be developed to extract the weak and the strong phases from the data.
According to the fitted result, the phenomenology analysis and discussions will be
present in Sec\ref{sec:Discussions}. Conclusion will be given in Sec\ref{sec:Conclusion}.

\section{Model\label{sec:Puzzle-and-method}}

\subsection{Parameterizations}

In general, the amplitudes for $B\to\pi K$ decays with $B=B^{\pm},B^{0}(\bar{B}^{0})$,
$K=K^{\pm},K^{0}(\bar{K}^{0})$, and $\pi=\pi^{\pm},\,\pi^{0}$, could
be parameterized as follows \citep{1126-6708-1999-02-014,2000NuPhB.569....3B,BENEKE2001245,Gronau:1998fn,2007IJMPA..22.2057I}
\begin{eqnarray}
A(\pi^{-}\bar{K}^{0}) & = & P^{\prime}(1+\epsilon_{a}e^{i\phi_{a}}e^{-i\gamma})\,,\label{eq:AP1}\\
-\sqrt{2}A(\pi^{0}K^{-}) & = & P^{\prime}(1+\epsilon_{a}e^{i\phi_{a}}e^{-i\gamma} \label{eq:AP2}\\
 &  & -\epsilon_{3/2}e^{i\phi}(e^{-i\gamma}-qe^{i\omega}))\,,\nonumber \\
-A(\pi^{+}K^{-}) & = & P^{\prime}(1+\epsilon_{a}e^{i\phi_{a}}e^{-i\gamma} \label{eq:AP3}\\
 &  & -\epsilon_{T}e^{i\phi_{T}}(e^{-i\gamma}-q_{C}e^{i\omega_{C}}))\,,\nonumber \\
\sqrt{2}A(\pi^{0}\bar{K}^{0}) & = & P^{\prime}(1+\epsilon_{a}e^{i\phi_{a}}e^{-i\gamma} \\
 &  & +(\epsilon_{3/2}e^{i\phi}-\epsilon_{T}e^{i\phi_{T}})e^{-i\gamma}\nonumber \label{eq:AP4}\\
 &  & +(\epsilon_{T}q_{C}e^{i(\phi_{T}+\omega_{C})}-\epsilon_{3/2}qe^{i(\phi+\omega)}))\,,\nonumber
\end{eqnarray}
where $A(\pi K)$ mean the amplitudes of $B\to\pi K$ decays. In the
above expressions, the ``unitarity triangle'' has been used to translate
the contributions from $\lambda_{t}$ to the $\lambda_{u}$ and $\lambda_{c}$
terms, where $\lambda_{q}=V_{qb}^{*}V_{qd}$ for $q=u,\,c,\,t$. $P^{\prime}$
denotes the major penguin amplitude (containing $\lambda_{c}$). The
weak angle $\gamma=arg(V_{ub}^{*})$ comes from the terms containing
$\lambda_{u}$. The others are real parameters, $\epsilon_{a}$,
$\epsilon_{3/2}$, $\epsilon_{T}$, $q$ and $q_{C}$ and their associated
strong phases $\phi_{a},\,\phi,\,\phi_{T},\,\omega,\,\omega_{C}$
\citep{BENEKE2001245}. The amplitudes for $B^{+,0}$ are obtained
by replacing the weak angle $\gamma$ by $-\gamma$. Some relative
minor terms such as annihilation, exchange and penguin-annihilation
terms have been neglected.

In Eqs.(\ref{eq:AP1}-\ref{eq:AP4}), the isospin symmetry is assumed. The initial
$B$ meson $(B^{-},\,\bar{B}^{0})$ is of $I_{B}=-\frac{1}{2}$ state
$|B(I=-\frac{1}{2})\rangle$. The final $(\pi K)$ state can be decomposed
into the $I_{\pi K}=+\frac{1}{2}$ state $\langle\pi(I=1)K(I=-\frac{1}{2})|$,
the $I_{\pi K}=-\frac{1}{2}$ state $\langle\pi(I=0)K(I=-\frac{1}{2})|$,
and the $I=-\frac{3}{2}$ state $\langle\pi(I=-1)K(I=-\frac{1}{2})|$.
The effective weak Hamiltonian has $\Delta I=0$ and $\Delta I=1$
operators, where $\Delta I\equiv|I_{B}-I_{\pi K}|$. Therefore, the
matrix element $\langle\pi K|H_{eff}|B\rangle$ could be described
by three isospin amplitudes $B_{1/2}$, $A_{1/2}$, and $A_{3/2}$,
which correspond to $\Delta I=0$ with $I_{\pi K}=-\frac{1}{2}$,
$\Delta I=1$ with $I_{\pi K}=+\frac{1}{2}$, $\Delta I=1$ with $I_{\pi K}=-\frac{3}{2}$,
respectively. The decay amplitudes are expressed as \citep{1126-6708-1999-02-014}
\begin{eqnarray}
A(\pi^{-}\bar{K}^{0}) & = &B_{1/2}+A_{1/2}+A_{3/2}\,,\\
-\sqrt{2}A(\pi^{0}K^{-}) & = & B_{1/2}+A_{1/2}-2A_{3/2}\,,\\
-A(\pi^{+}K^{-}) & = & B_{1/2}-A_{1/2}-A_{3/2}\,,\\
\sqrt{2}A(\pi^{0}\bar{K}^{0}) & = & B_{1/2}-A_{1/2}+2A_{3/2}\,,
\end{eqnarray}
from which the quadrangle relation \citep{Nir:1991cu,Gronau:1991dq}
is obtained
\begin{eqnarray}
 &  & A(\pi^{-}\bar{K}^{0})+\sqrt{2}A(\pi^{0}K^{-})\label{eq:quadrangle relation}\\
 & = & A(\pi^{+}K^{-})+\sqrt{2}A(\pi^{0}\bar{K}^{0})\nonumber \\
 & = & 3A_{3/2}\,.\nonumber
\end{eqnarray}
These isospin amplitudes $B_{1/2},\,A_{1/2},\,A_{3/2}$ need to be
process-independent. The same argument is applied to the eleven parameters,
$P^{\prime}$, $\epsilon_{a}$, $\epsilon_{3/2}$, $\epsilon_{T}$,
$q$, $q_{C}$, $\phi_{a},\,\phi,\,\phi_{T},\,\omega,\,\omega_{C}$.
One should note that the isospin symmetry and the process-independence
are closely correlated.
It is interesting to examine
whether the isospin symmetry could be preserved or broken,
if the parameters are process-dependent.
This will be made in latter text.

There are total twelve parameters to be determined, but we have only
eight independent measurements, four sets of branching rates and asymmetries.
(We identify the measurement for the mixing induced CP asymmetry $S_{\pi^{0}K_{S}^{0}}$
as an independent test for the weak angle $\beta$ and our model.)
It is impossible to completely determine all the parameters.
In the next section, the QCDF with three parton corrections is used to calculate
the eleven hadronic parameters.
Further more, the three phases $\phi_{a},\,\phi,\,\phi_{T}$ are assumed to contain
both perturbative and nonperturbative QCD contributions.
The perturbative parts $\hat{\phi}_{a},\,\hat{\phi},\,\hat{\phi}_{T}$, are calculated by the QCDF formalism.
The nonperturbative $\bar{\phi}_{a},\,\bar{\phi},\,\bar{\phi}_{T}$ are determined by a least squares fit to the experimental data.
We will also assume that the factorization scale $\mu$ for four processes are different,
$\mu_{i}$, $i=1,\cdots,4$.
Then, there are total eight parameters, $\bar{\phi}_{a},\,\bar{\phi},\,\bar{\phi}_{T}$, $\gamma$, and $\mu_{i}$,
which can be completely determined by eight measurements.
We identify this model as the minimal QCDF model (MQCDF).

\subsection{Observables}

In order to explicitly explore how the puzzle would happen, every
term is remained in the following expressions without applying any
approximation. Using the above parametrization for the amplitudes, the branching ratios
are expressed as

\begin{eqnarray}
Br(\pi^{-}K^{0}) & = & \Gamma_{B^{-}}P^{\prime2}[1+2\epsilon_{a}\cos(\phi_{a})\cos(\gamma)+\epsilon_{a}^{2}]\,,\label{eq:Br1}
\end{eqnarray}
\begin{align}
Br(\pi^{0}K^{-}) & =\frac{1}{2}\Gamma_{B^{-}}P^{\prime2}[1+\epsilon_{3/2}^{2}+\epsilon_{a}{}^{2}+\epsilon_{3/2}^{2}q^{2}\label{eq:Br2}\\
 & +2\epsilon_{a}\cos(\phi_{a})\cos(\gamma)-2\epsilon_{3/2}\text{cos}(\phi)\cos(\gamma)\nonumber \\
 & +2\epsilon_{3/2}q\text{cos}(\omega+\phi)-2\epsilon_{3/2}\epsilon_{a}\text{cos}(\phi-\phi_{a})\nonumber \\
 & -2\epsilon_{3/2}^{2}q\text{cos}(\omega)\cos(\gamma)\nonumber \\
 & +2\epsilon_{3/2}\epsilon_{a}q\text{cos}(\omega+\phi-\phi_{a})\cos(\gamma)]\,,\nonumber
\end{align}
\begin{align}
Br(\pi^{+}K^{-}) & =\Gamma_{B^{0}}P^{\prime2}[1+\epsilon_{T}^{2}+\epsilon_{a}{}^{2}+\epsilon_{T}^{2}q^{2}\label{eq:Br3}\\
 & +2\epsilon_{a}\cos(\phi_{a})\cos(\gamma)-2\epsilon_{T}\text{cos}(\phi_{T})\cos(\gamma)\nonumber \\
 & +2\epsilon_{T}q\text{cos}(\omega+\phi_{T})-2\epsilon_{T}\epsilon_{a}\text{cos}(\phi_{T}-\phi_{a})\nonumber \\
 & -2\epsilon_{T}^{2}q\text{cos}(\omega)\cos(\gamma)\nonumber \\
 & +2\epsilon_{T}\epsilon_{a}q\text{cos}(\omega+\phi_{T}-\phi_{a})\cos(\gamma)]\,,\nonumber
\end{align}
\begin{align}
Br(\pi^{0}K^{0}) & =\frac{1}{2}\Gamma_{B^{0}}P^{\prime2}[1+\epsilon_{a}{}^{2}+\epsilon_{3/2}^{2}\label{eq:Br4}\\
 & +\epsilon_{3/2}^{2}q^{2}+\epsilon_{T}^{2}+\epsilon_{T}^{2}q_{C}^{2}\nonumber \\
 & +2\epsilon_{a}\cos(\phi_{a})\cos(\gamma)-2\epsilon_{3/2}\text{cos}(\phi)\cos(\gamma)\nonumber \\
 & -2\epsilon_{T}\text{cos}(\phi_{T})\cos(\gamma)+2\epsilon_{3/2}q\text{cos}(\omega+\phi)\nonumber \\
 & -2\epsilon_{3/2}\epsilon_{a}\text{cos}(\phi-\phi_{a})+2\epsilon_{T}q_{C}\text{cos}(\omega_{C}+\phi_{T})\nonumber \\
 & -2\epsilon_{T}\epsilon_{a}\text{cos}(\phi_{T}-\phi_{a})-2\epsilon_{3/2}^{2}q\text{cos}(\omega)\cos(\gamma)\nonumber \\
 & -2\epsilon_{T}^{2}q_{C}\text{cos}(\omega_{C})\cos(\gamma)\nonumber \\
 & +2\epsilon_{3/2}\epsilon_{a}q\text{cos}(\omega+\phi-\phi_{a})\cos(\gamma)\nonumber \\
 & +2\epsilon_{T}\epsilon_{a}q_{C}\text{cos}(\omega_{C}+\phi_{T}-\phi_{a})\cos(\gamma)]\,,\nonumber
\end{align}
and the direct CP asymmetries written as
\begin{eqnarray}
A_{CP}(\pi^{-}K^{0}) & = & N(\pi^{-}K^{0})\epsilon_{a}\sin(\phi_{a})\sin(\gamma)\,,\label{eq:ACP1}\\
A_{CP}(\pi^{0}K^{-}) & = & N(\pi^{0}K^{-})\sin(\gamma)[\epsilon_{a}\sin(\phi_{a})\label{eq:ACP2}\\
 &  &-\epsilon_{3/2}\sin(\phi) +\epsilon_{3/2}^{2}q\sin(\omega)\nonumber\\
 &  &-\epsilon_{a}\epsilon_{3/2}q\sin(-\phi_{a}+\phi+\omega)]\,,\nonumber \\
A_{CP}(\pi^{+}K^{-}) & = & N(\pi^{+}K^{-})\sin(\gamma)[\epsilon_{a}\sin(\phi_{a})\label{eq:ACP3}\\
 &  & -\epsilon_{T}\sin(\phi_{T})+\epsilon_{T}^{2}q_{C}\sin(\omega_{C})\nonumber\\
 &  & -\epsilon_{a}\epsilon_{T}q\sin(-\phi_{a}+\phi_{T}+\omega_{C})]\,,\nonumber \\
A_{CP}(\pi^{0}K^{0}) & = & N(\pi^{0}K^{0})\sin(\gamma)[\epsilon_{a}\sin(\phi_{a})\label{eq:ACP4}\\
 &  & +(\epsilon_{3/2}\sin(\phi)-\epsilon_{T}\sin(\phi_{T}))\nonumber \\
 &  & +(\epsilon_{3/2}^{2}q\sin(\omega)+\epsilon_{T}^{2}q_{C}\sin(\omega_{C}))\nonumber \\
 &  & +\epsilon_{a}(\epsilon_{3/2}q\sin(-\phi_{a}+\phi+\omega)\nonumber \\
 &  & -\epsilon_{T}q_{C}\sin(-\phi_{a}+\phi_{T}+\omega_{C}))\nonumber \\
 &  & -\epsilon_{3/2}(\epsilon_{T}q\sin(\phi-\phi_{T}+\omega)\nonumber \\
 &  & +\epsilon_{T}q_{C}\sin(-\phi+\phi_{T}+\omega_{C}))]\,,\nonumber
\end{eqnarray}
where
\begin{eqnarray*}
N(\pi K) & = & \frac{2P^{^{\prime}2}\Gamma_{B}}{Br(\pi K)}\,.
\end{eqnarray*}
The branching ratios are calculated according to
\begin{eqnarray*}
Br(B\to\pi K) & = & \frac{1}{2}\Gamma_{B}(\left|A(B\to\pi K)\right|^{2}+\left|A(\bar{B}\to\pi K)\right|^{2})
\end{eqnarray*}
where
\begin{eqnarray*}
\Gamma_{B} & = & \frac{\tau_{B}}{16\pi m_{B}}\,.
\end{eqnarray*}
According to the particle data group (PDG) definition \citep{Patrignani:2016xqp},
the CP asymmetry is defined by
\begin{eqnarray*}
A_{CP} & = & \frac{Br(\bar{B}\to\bar{f})-Br(B\to f)}{Br(\bar{B}\to\bar{f})+Br(B\to f)}\,.
\end{eqnarray*}
Three ratios
\begin{eqnarray}
R & = & \frac{Br(\pi^{+}K^{-})+Br(\pi^{0}K^{0})}{Br(\pi^{-}K^{0})+Br(\pi^{0}K^{-})}\frac{\Gamma_{B^{-}}}{\Gamma_{B^{0}}}\,,\label{eq:R}
\end{eqnarray}
\begin{eqnarray}
R_{c} & = & 2\left[\frac{Br(B^{-}\to\pi^{0}K^{-})}{Br(B^{-}\to\pi^{-}\bar{K}^{0})}\right]\,,\label{eq:Rc}
\end{eqnarray}
\begin{eqnarray}
R_{n} & = & \frac{1}{2}\left[\frac{Br(\bar{B}^{0}\to\pi^{-}K^{+})}{Br(\bar{B}^{0}\to\pi^{0}K^{0})}\right]\,,\label{eq:Rn}
\end{eqnarray}
have been widely used in literature as tests for the SM. The comparisons
between the theoretical predictions and experimental data of these
quantities are left to Sec.\ref{sec:Discussions}.

\section{Calculations\label{sec:Calculations}}

At the factorization scale $\mu\sim m_{b}$ (the bottom quark
mass), the effective weak Hamiltonian for $B\to\pi K$ decays is given
by \citep{Buchalla:1995vs,Buras:1998raa}

\begin{eqnarray}
H_{eff} & = & \frac{G_{F}}{\sqrt{2}}\sum_{p=u,c}\lambda_{p}\left[C_{1}(\mu)Q_{1}^{p}(\mu)+C_{2}(\mu)Q_{2}^{p}(\mu)\right.\label{eq:Eff H}\\
 &  & +\sum_{i=3,\cdots,10}C_{i}(\mu)Q_{i}(\mu)+C_{7\gamma}(\mu)Q_{7\gamma}(\mu)\nonumber \\
 &  & \left.+C_{8G}(\mu)Q_{8G}(\mu)\right]+h.c.\nonumber
\end{eqnarray}
where $\lambda_{p}=V_{ps}^{*}V_{pb}$ is the product of CKM matrix
elements. The local $\Delta B=1$ four quark operators $Q_{i}$ are
defined as
\begin{eqnarray}
Q_{1}^{p} & = & (\bar{s}_{\alpha}p_{\alpha})_{(V-A)}(\bar{p}_{\beta}b_{\beta})_{(V-A)}\,,\\
Q_{2}^{p} & = & (\bar{s}_{\alpha}p_{\beta})_{(V-A)}(\bar{p}_{\beta}b_{\alpha})_{(V-A)}\,,\nonumber \\
Q_{3,5} & = & (\bar{s}_{\beta}b_{\beta})_{(V-A)}\sum_{q^{\prime}}(\bar{q}_{\alpha}^{\prime}q_{\alpha}^{\prime})_{(V\mp A)}\,,\nonumber \\
Q_{4,6} & = & (\bar{s}_{\beta}b_{\alpha})_{(V-A)}\sum_{q^{\prime}}(\bar{q}_{\alpha}^{\prime}q_{\beta}^{\prime})_{(V\mp A)}\,,\nonumber \\
Q_{7,9} & = & \frac{3}{2}(\bar{s}_{\beta}b_{\beta})_{(V-A)}\sum_{q^{\prime}}e_{q^{\prime}}(\bar{q}_{\alpha}^{\prime}q_{\alpha}^{\prime})_{(V\pm A)}\,,\nonumber \\
Q_{8,10} & = & \frac{3}{2}(\bar{s}_{\beta}b_{\alpha})_{(V-A)}\sum_{q^{\prime}}e_{q^{\prime}}(\bar{q}_{\alpha}^{\prime}q_{\beta}^{\prime})_{(V\mp A)}\,,\nonumber \\
Q_{7\gamma} & = & -\frac{e}{8\pi^{2}}(\bar{q}\sigma^{\mu\nu}(1+\gamma_{5})b)F_{\mu\nu}\,,\nonumber \\
Q_{8G} & = & -\frac{g}{8\pi^{2}}m_{b}(\bar{q}\sigma^{\mu\nu}(1+\gamma_{5})b)G_{\mu\nu}\,,\nonumber
\end{eqnarray}
where $q^{\prime}\in\left\{ u,d,s,c,b\right\} $, $\alpha$ and $\beta$
mean color indices, and $e_{q^{\prime}}=2/3(-1/3)$ for $u(d)$ type
quarks. The Wilson coefficients $C_{i}(\mu)$ collect the radiative
contributions between $\mu$ and $M_{W}$ up to next-to-leading order
(NLO) in $\alpha_{s}$. The renormalization scheme is chosen as the
minimal-subtraction ($\overline{MS}$) with $\Lambda_{\overline{MS}}^{(5)}=0.225$
GeV. The $C_{i}$ are calculated by the naive dimensional regularization
(NDR).

Assume the naive factorization $\langle\pi K|Q_{i}(0)|B\rangle=\langle\pi|\bar{q}_{1}\Gamma_{i,\mu}b|B\rangle\langle K|\bar{q}_{2}\Gamma_{i}^{\mu}q_{3}|0\rangle$
or $\langle K|\bar{q}_{1}\Gamma_{i,\mu}b|B\rangle\langle\pi|\bar{q}_{2}\Gamma_{i}^{\mu}q_{3}|0\rangle$,
with $\Gamma_{u,i}\otimes\Gamma_{i}^{\mu}=\gamma_{\mu}(1-\gamma_{5})\otimes\gamma^{\mu}(1\pm\gamma_{5})$,
and include possible one loop radiative corrections to the factorized
terms, the hadronic matrix element $\langle\pi K|Q_{i}(0)|B\rangle$
could be expressed in the following factorized form, up-to the twist-3
and NLO in $\alpha_{s}$,
\begin{eqnarray}
 &  & \langle\pi K|Q_{i}(0)|B\rangle\label{eq:fact-formula}\\
 & = & F_{0}^{B\pi}(0)\int_{0}^{1}du\text{Tr}[T_{i,K}^{I}(u)\phi^{K}(u)]\nonumber \\
 &  & +F_{0}^{BK}(0)\int_{0}^{1}du\text{Tr}[T_{i,\pi}^{I}(u)\phi^{\pi}(u)]\nonumber \\
 &  & +\int_{0}^{\infty}d\xi\int_{0}^{1}du\int_{0}^{1}dv\text{Tr}[\phi^{B}(\xi)T_{i,B\pi K}^{II}(\xi,\,u,\,v)\nonumber\\
 &&    \phi^{\pi}(u)\phi^{K}(v)]\nonumber
\end{eqnarray}
where $\text{Tr}$ means the trace taken over the spin indices and
the integrals are made over the momentum fractions $u,\,v$, $\xi$.
The hard scattering kernels, $T^{I,II}$, describe the short distance
interactions between partons of the external initial and final state
mesons. The kernel $T^{I}$ contains tree (T), vertex (V), and penguin
(P) contributions. The kernel $T^{II}$ contains hard spectator (HS).
The transition form factors, $F_{0}^{B\pi}$ and
$F_{0}^{BK}$, and the meson spin distribution amplitudes $\phi^{B}$,
$\phi^{\pi}$, $\phi^{K}$ encode the long distance interactions of
the quarks and gluons. The validity of the factorization formula is
explained below.

The factorization at leading twist order has been well-known \citep{BENEKE2001245,2003NuPhB.675..333B}.
In the following, let's concentrate on the twist-3 level.

At the twist-3 two parton order, the hard spectator terms may contain
end-point divergences in the form
\begin{eqnarray}
X_{H} & = & \int_{0}^{1}du\frac{\phi_{p}(u)}{u}\,,
\end{eqnarray}
 if the twist-3 two parton pseudo-scalar distribution amplitude $\phi_{p}$
is a constant \citep{2003NuPhB.675..333B,BENEKE2001245}. This viewpoint
of a constant $\phi_{p}$ has been widely employed in literature.
However, this is not the correct fact, because a constant $\phi_{p}$
is determined by the equation of motion (EOM) at the chiral symmetry
breaking scale $\mu_{c}\sim1$ GeV. The partons involving in the hard
scattering kernels $T^{II}$ would have energies about the energetic
scale $\mu_{E}\sim m_{b}\gg\mu_{c}$. The correct EOM of $\phi_{p}$
would be taken at the energetic scale $\mu_{E}$ instead of $\mu_{c}$.
As a result, $\phi_{p}$ becomes equal to $\phi_{\sigma}$ and not
a constant. The above mentioned end-point divergences in the hard
spectator terms would vanish \citep{Yeh:2008xc}.

There also exist end-point divergences from the annihilation terms
in the form
\begin{eqnarray}
X_{A} & = & \int_{0}^{1}du\frac{\phi_{P}(u)}{u(u-\xi)}\,,
\end{eqnarray}
even for the twist-2 pseudo-scalar distribution amplitude $\phi_{P}$
\citep{BENEKE2001245,2003NuPhB.675..333B}. The regularization method
is not to neglect the momentum fraction factor $\xi$ from the spectator
lines of the $B$ meson. It can be shown that the regularized result
is equivalent to include the twist-4 contributions \citep{Yeh:2007fi}.
As a result, the annihilation terms up to twist-3 are also factorizable.
The annihilation terms are not included in the above factorization
formula and will be neglected in later calculations.

According to the study by Yeh \citep{Yeh:2007fi}, there are five
types of different ways as shown in Fig.1 that the three parton state
$|q\bar{q}g\rangle$ of the emitted final state meson $M=\pi,\,K$
can contribute at one loop level. The three parton state from the
$B$ meson needs not be considered, because only soft spectator gluonic
partons can involve and their contributions are power suppressed by
$O(1/m_{B}^{2})$. By power counting \citep{Yeh:2008xc}, it is easy
to see that only contributions from Fig. 1(a) are leading at twist-3
order (i.e., suppressed by $O(1/m_{B})$) and the other types
of contributions are at least of twist-4 (i.e., suppressed by $O(1/m_{B}^{2})$
than the leading twist-2 term). Considering only contributions from
Fig.1(a), the meson spin distribution amplitude $\phi^{M}(u)$ for
$M=\pi,\,K$ has the following spin representation \citep{Yeh:2008xc}
\begin{align}
 & \phi^{M}(u)\\
= & \int_{0}^{\infty}\frac{d\lambda}{2\pi}e^{-iu\lambda}\langle M|\bar{q}_{a}(0)q_{b}(\frac{\lambda}{E}n)|0\rangle\nonumber \\
 & +\int_{0}^{\infty}\frac{d\lambda}{2\pi}\int_{0}^{\infty}\frac{d\eta}{2\pi}e^{-i\alpha\lambda}e^{-i\beta\eta}\langle M_{2}|\bar{q}_{a}(0)ig\not\!A(\frac{\eta}{E}n)q_{b}(\frac{\lambda}{E}n)|0\rangle\nonumber \\
= & -\frac{if_{M}}{4N_{c}}\left[\gamma_{5}\not q\phi_{P}(u)+\mu_{\chi}\left(\gamma_{5}\phi_{p}(u)-\frac{1}{2}\epsilon_{\perp}\cdot\sigma\phi_{\sigma}(u)\right.\right.\nonumber \\
 & \left.\left.+2\gamma_{5}\delta(u-\alpha-\beta)\int_{0}^{1}d\alpha\int_{0}^{(1-\alpha)}d\beta\frac{\phi_{3p}(\alpha,\beta)}{\alpha\beta}\right)\right]\nonumber
\end{align}
where $\epsilon_{\perp}\cdot\sigma=\epsilon_{\alpha\beta\eta\lambda}\sigma^{\alpha\beta}\bar{n}^{\gamma}n^{\lambda}$
with $\bar{n}^{\mu}$ and $n^{\mu}$ being light like unit vector satisfying $\bar{n}^2=n^2=0$ and $\bar{n}\cdot n=1$.
$f_{M}$ is the decay constant and $\mu_{\chi}^{M}=m_{M}^{2}/(m_{q_{a}}+m_{q_{b}})$
is the chiral enhanced factor. $\phi_{P}$ is the twist-2 pseudo-scalar
distribution function and $\phi_{p}$ and $\phi_{\sigma}$ are twist-3
two parton pseudo-scalar and pseudo-tensor distribution functions.
The three parton distribution function $\phi_{3p}$ is defined by
\begin{align}
 & \int_{0}^{\infty}\frac{d\lambda}{2\pi}\int_{0}^{\infty}\frac{d\eta}{2\pi}e^{-i\alpha\lambda}e^{-i\beta\eta}\\
 & \langle M_{2}|\bar{q}(0)\gamma_{5}\sigma^{\alpha\beta}gG^{\mu\nu}(\frac{\eta}{E}n)q(\frac{\lambda}{E}n)|0\rangle\nonumber \\
= & -if_{M_{2}}\mu_{\chi}P^{\alpha\beta\mu\nu}(q)\phi_{3p}(\alpha,\beta)\nonumber
\end{align}
and has a parametrization
\begin{eqnarray}
\phi_{3p}(\alpha,\beta) & = & 360\eta\alpha\beta^{2}(1-\alpha-\beta)(1+\frac{1}{2}\omega(7\beta-3))
\end{eqnarray}
with $\eta=0.015$ and $\omega=-3$. This representation of the spin
distribution $\phi^{M}$ is written according to the following facts:
\begin{itemize}
\item The effective spin structure and the integrals of $\phi_{3p}$
can be derived from the tree level Feynman diagrams similar to
Fig.1(a), in which the radiative gluon is absorbed by the vertex.
The detailed derivations refer to \citep{Yeh:2008xc}. When the three
parton one loop Feynman diagrams as depicted in Fig.1(a) are considered,
it can be easily derived that the same spin structure is also applicable.
Effectively, one may regard the three parton in this scenario as a
pseudo-scalar two parton term. This shows a very convenient way to
calculate the three parton one loop corrections by referring
to those one loop calculations of $\phi_{p}$ \citep{Yeh:2007fi}.
\item It can be shown that the energetic EOM $\phi_{p}=\phi_{\sigma}$ \citep{Yeh:2008xc}
would not change if $\phi_{3p}$ is considered. This is because the
spin projector $P^{\alpha\beta\mu\nu}(q)$ vanishes at the energetic
EOM condition. It implies that $\phi_{p}$ and $\phi_{\sigma}$ decouple
from $\phi_{3p}$ at the energetic limit $E\gg\Lambda_{QCD}$.
\end{itemize}
Combining the above two facts, it can be shown that the one loop corrections
of $\phi_{3p}$ are factorizable according to the analysis for the
factorizability of the one loop corrections of $\phi_{p}$ given in
\citep{Yeh:2007fi}. The complete analysis of the above arguments
will be present elsewhere. The $B$ meson spin distribution amplitude
is given by \citep{Yeh:2007fi}
\begin{eqnarray}
 & \phi^{B}(\xi) & =\int_{0}^{\infty}\frac{d\lambda}{2\pi}e^{-i\xi\lambda}\langle0|\bar{q}(0)b(\frac{\lambda}{E}n)|B\rangle\\
 &  & =\frac{if_{B}}{4N_{c}}\left[(\not P_{B}+m_{B})\gamma_{5}\phi_{B}(\xi)\right]\,.\nonumber
\end{eqnarray}
For numeric calculations, we employ the following models
\begin{eqnarray}
\phi_{P}(u) & = & \phi_{p}(u)=6u(1-u)\,,\\
\phi_{B}(\xi) & = & \frac{N_{B}\xi^{2}\bar{\xi}^{2}}{[\xi^{2}+\epsilon_{B}\bar{\xi}]^{2}}\,,
\end{eqnarray}
with $N_{B}=0.133$, $\epsilon_{B}=0.005$ for $\lambda_{B}=350\;\text{MeV}$.
$N_{B}$ and $\epsilon_{B}$ are determined by the moments
\begin{eqnarray}
\int_{0}^{1}d\xi\phi_{B}(\xi) & = & 1,\;\int_{0}^{1}d\xi\frac{\phi_{B}(\xi)}{\xi}=\frac{m_{B}}{\lambda_{B}}\,,
\end{eqnarray}
where the errors are controlled within $1\%$. Since the radiative
corrections for the meson distribution amplitudes start at NNLO
$O(\alpha_{s}^{2})$, they could be neglected at NLO calculations.

Summarizing the above analysis, it shows that the factorization formula
of QCDF given in Eq.(\ref{eq:fact-formula}) is valid up-to complete
twist-3 power corrections and NLO in $\alpha_{s}$. Due to their smallness
as compared with the other types of contributions, we will completely
neglect the annihilation terms. This makes our later explanations
for the puzzle completely different from most literature based on
QCDF, in which the annihilation electro-weak penguin terms would be
important \citep{BENEKE2001245,2003NuPhB.675..333B}.

In the following, the tree, vertex, penguin, and hard spectator contributions
are collected as $a^{p}_{i}(\pi K)$ for each $Q_{i}$ as given below
\begin{eqnarray}
 &  & a_{i}^{p}(\pi K,\mu)\label{eq:ai-expansion}\\
 & = & (C_{i}(\mu)+\frac{C_{i\pm1}(\mu)}{N_{c}})N_{i}(K)\nonumber \\
 & + & \frac{C_{i\pm1}(\mu)}{N_{c}}\frac{\alpha_{s}(\mu)C_{F}}{4\pi}\left[V_{i}^{(2)}(K)+V_{i}^{(3)}(K)\right]\nonumber \\
 & + & (4\pi\alpha_{s}(\mu)C_{F})\frac{C_{i\pm1}(\mu)}{N_{c}^{2}}\left[H_{i}^{(2)}(\pi K)+H_{i}^{(3)}(\pi K)\right]\nonumber \\
 & + & P_{i}^{p(2)}(K,\mu)+P_{i}^{p(3)}(K,\mu)\,.\nonumber
\end{eqnarray}

The expressions for the tree $N_{i}(K)$, the two parton vertex $V_{i}^{(2)}(K)$,
the two parton hard spectator $H_{i}^{(2)}(\pi K)$, and the two
parton penguin $P_{i}^{p(2)}(K)$ are referred to \citep{Yeh:2007fi,BENEKE2001245}.
Here, we only present the three parton vertex $V_{6,8}^{(3)}(K)$,
the three parton hard spectator $H_{i}^{(3)}(\pi K)$,
and the three parton penguin $P_{6,8}^{p(3)}(K,\mu)$ as follows

\begin{eqnarray}
 & V_{6,8}^{(3)}(K)\\
= & 2\int_{0}^{1}d\alpha\int_{0}^{(1-\alpha)}d\beta & {\displaystyle \left[\frac{\phi_{3p}(\alpha,\beta)(-6+h^{(3)}(\alpha,\beta))}{\alpha\beta}\right]}\,,\nonumber
\end{eqnarray}
where the kernel is
\begin{eqnarray}
h^{(3)}(\alpha,\beta) & = & 2\left[\text{Li}_{2}(x)+\frac{1}{2}(\ln\bar{x})^{2}-(x\leftrightarrow(1-x))\right]_{x=\alpha+\beta}\,.
\end{eqnarray}
 The penguin functions $P_{6,8}^{p(3)}(K,\,\mu)$ are given by
\begin{eqnarray}
 &  & P_{6}^{p(3)}(K,\mu)\\
 & = & \frac{\alpha_{s}(\mu)C_{F}}{4\pi N_{c}}\left\{ C_{1}(\mu)\left[\left(\frac{4}{3}\ln\frac{m_{b}}{\mu}+\frac{2}{3}\right)A_{G}-G_{K}^{(3)}(s_{p})\right]\right.\nonumber \\
 &  & +C_{3}(\mu)\left[2\left(\frac{4}{3}\ln\frac{m_{b}}{\mu}+\frac{2}{3}\right)A_{G}-G_{K}^{(3)}(0)-G_{K}^{(3)}(1)\right]\nonumber \\
 &  & +(C_{4}(\mu)+C_{6}(\mu))\left[\frac{4}{3}n_{f}\left(\ln\frac{m_{b}}{\mu}\right)A_{G}\right.\nonumber \\
 &  &\left.-(n_{f}-2)G_{K}^{(3)}(0)-G_{K}^{(3)}(s_{c})-G_{K}^{(3)}(1)\right]\nonumber \\
 &  & \left.-4C_{8g}^{\text{eff}}\int_{0}^{1}d\alpha\int_{0}^{(1-\alpha)}d\beta\frac{\phi_{3p}(\alpha,\beta)}{\alpha\beta}\right\} \,,\nonumber \\
 &  & P_{8}^{p(3)}(\mu)\\
 & = & \frac{\alpha}{9\pi N_{c}}\left\{ (C_{1}(\mu)+N_{c}C_{2}(\mu))\right.\nonumber \\
 &  & \left[\frac{4}{3}A_{G}\ln\frac{m_{b}}{\mu}+\frac{2}{3}A_{G}-G_{K}^{(3)}(s_{p})\right]\nonumber \\
 &  & \left.-6C_{7\gamma}^{\text{eff}}\int_{0}^{1}d\alpha\int_{0}^{(1-\alpha)}d\beta\frac{\phi_{3p}(\alpha,\beta)}{\alpha\beta}\right\} \,,\nonumber
\end{eqnarray}
where $n_{f}=5$ denotes the number of the flavors.

The functions $G_{K}^{(3)}(s_{p})$ is defined as
\begin{eqnarray*}
\hat{G}_{K}^{(3)}(s_{p}) & = & 2\int_{0}^{1}d\alpha\int_{0}^{(1-\alpha)}d\beta\frac{\phi_{3p}(\alpha,\beta)}{\alpha\beta}G(s-i\epsilon,1-x)|_{x=\alpha+\beta}\,,\\
G(s,u) & = & -4\int_{0}^{1}dxx(1-x)\ln[s-x(1-x)u]\,.
\end{eqnarray*}
By substituting $s_{p}=0,\,m_{c}^{2}/m_{b}^{2},\,1$,
$\hat{G}_{K}^{(3)}(s_{p})$ are calculated for later numerical analysis
\begin{eqnarray*}
\hat{G}_{K}^{(3)}(0) & = & 0.81-i1.23\,,\\
\hat{G}_{K}^{(3)}(s_{c}) & = & 1.26-i0.94\,,\\
\hat{G}_{K}^{(3)}(1) & = & 0.059\,.
\end{eqnarray*}
The effective Wilson coefficients $C_{7\gamma}^{\text{eff}}$ and
$C_{8g}^{\text{eff}}$ are calculated at their leading order in $\alpha_{s}$
to be constants
\begin{eqnarray}
C_{7\gamma}^{\text{eff}} & = & -0.32\,,\\
C_{8g}^{\text{eff}} & = & -0.15\,.
\end{eqnarray}
The hard spectator functions $H_{i}^{(3)}(\pi K)$ are, for $i=1-5,\,7,\,9,\,10$,
\begin{eqnarray}
 &  & H_{i}^{(3)}(\pi K)\\
 & = & 2\frac{B_{\pi K}}{A_{\pi K}}m_{B}r_{\chi}^{\pi}(\mu_{h})\int_{0}^{1}d\xi\frac{\phi_{B}(\xi)}{\xi}\int_{0}^{1}d\alpha\int_{0}^{(1-\alpha)}d\beta\int_{0}^{1}du\nonumber\\
 & \times & \left[\frac{\phi_{3p}(\alpha,\beta)\phi_{P}(u)}{(\alpha+\beta)\alpha\beta(\bar{u}-\xi)}\right]\,,\nonumber
\end{eqnarray}
where
\begin{eqnarray}
B_{\pi K} & = & i\frac{G_{F}}{\sqrt{2}}f_{B}f_{K}f_{\pi}
\end{eqnarray}
and $H_{6,8}^{(2,3)}(\pi K)=0$. The scale in $r_{\chi}^{\pi}(\mu_{h})$
is assumed as $\mu_{h}=\sqrt{\Lambda_{h}\mu}$ and $\Lambda_{h}=0.5\;\text{GeV}$.

According to the QCD factorization formula Eq.(1), we obtain the expressions
for the amplitudes \citep{2003NuPhB.675..333B,BENEKE2001245,Yeh:2007fi}

\begin{align}
A(\pi^{-}\bar{K}^{0}) & =\lambda_{p}A_{\pi K}\left[\left(a_{4}^{p}-\frac{1}{2}a_{10}^{p}\right)\right.\label{eq:QCDF-amp1}\\
 & \left.+r_{\chi}^{K}\left(a_{6}^{p}-\frac{1}{2}a_{8}^{p}\right)\right]\,,\nonumber \\
-\sqrt{2}A(\pi^{0}K^{-}) & =A_{\pi K}\left[\lambda_{u}a_{1}+\lambda_{p}\left(a_{4}^{p}+a_{10}^{p}\right)\right.\label{eq:QCDF-amp2}\\
 & \left.+\lambda_{p}r_{\chi}^{K}\left(a_{6}^{p}+a_{8}^{p}\right)\right]\nonumber \\
 & +A_{K\pi}\left[\lambda_{u}a_{2}+\lambda_{p}\frac{3}{2}\left(-a_{7}+a_{9}\right)\right]\,,\nonumber \\
-A(\pi^{+}K^{-}) & =A_{\pi K}\left[\lambda_{u}a_{1}+\lambda_{p}\left(a_{4}^{p}+a_{10}^{p}\right)\right.\label{eq:QCDF-amp3}\\
 & \left.+\lambda_{p}r_{\chi}^{K}\left(a_{6}^{p}+a_{8}^{p}\right)\right]\,,\nonumber \\
\sqrt{2}A(\pi^{0}\bar{K}^{0}) & =\lambda_{p}A_{\pi K}\left[\left(a_{4}^{p}-\frac{1}{2}a_{10}^{p}\right)\right.\label{eq:QCDF-amp4}\\
 & \left.+r_{\chi}^{K}\left(a_{6}^{p}-\frac{1}{2}a_{8}^{p}\right)\right]\nonumber \\
 & -A_{K\pi}\left[\lambda_{u}a_{2}+\lambda_{p}\frac{3}{2}\left(-a_{7}+a_{9}\right)\right]\,,\nonumber
\end{align}
where the $\mu$ dependence of $a_{i}$ is implicitly understood.
We have defined the following factors
\begin{eqnarray}
A_{\pi K} & = & i\frac{G_{F}}{\sqrt{2}}(m_{B}^{2}-m_{\pi}^{2})F_{0}^{B\pi}(m_{K}^{2})f_{K}\,,\\
A_{K\pi} & = & i\frac{G_{F}}{\sqrt{2}}(m_{B}^{2}-m_{K}^{2})F_{0}^{BK}(m_{\pi}^{2})f_{\pi}\,,\\
r_{\chi}^{K} & = & \frac{2m_{K}^{2}}{m_{b}(m_{q}+m_{s})}(1+A_{G})\,,\\
\lambda_{u}&=&\lambda_{c}\tan^{2}\theta_{c}R_{b}e^{-i\gamma}\,,\\
\tan^{2}\theta_{c} & = & \frac{\lambda^{2}}{1-\lambda^{2}}\,,\\
R_{b}&=&\frac{1-\lambda^{2}/2}{\lambda}\left|\frac{V_{ub}}{V_{cb}}\right|\,,\\
\lambda &=& |V_{us}|\,.
\end{eqnarray}
The three parton tree factor $A_{G}$ comes from the integration \citep{Yeh:2008xc}
\begin{eqnarray}
A_{G} & = & 2\int_{0}^{1}d\alpha\int_{0}^{(1-\alpha)}d\beta\frac{\phi_{3p}(\alpha,\beta)}{\alpha\beta}\,.
\end{eqnarray}

By comparing expressions given in Eqs.(\ref{eq:AP1}-\ref{eq:AP4}) and Eqs.(\ref{eq:QCDF-amp1}-\ref{eq:QCDF-amp4}),
the perturbative parts of the amplitude parameters are calculated
to be \citep{2003NuPhB.675..333B}
\begin{eqnarray}
P^{\prime} & = & \lambda_{c}A_{\pi K}\left[\left( a_{4}^{c}-\frac{1}{2}a_{10}^{c}\right)
                 +r_{\chi}^{K}\left(a_{6}^{c}-\frac{1}{2}a_{8}^{c}\right)\right]\,,\label{eq:QCDF-P}\\
\epsilon_{a}e^{i\hat{\phi}_{a}} & = & \epsilon_{KM}\frac{\left(a_{4}^{u}-\frac{1}{2}a_{10}^{u}\right)+r_{\chi}^{K}\left(a_{6}^{u}-\frac{1}{2}a_{8}^{u}\right)}{\left(a_{4}^{c}-\frac{1}{2}a_{10}^{c}\right)+r_{\chi}^{K}\left(a_{6}^{c}-\frac{1}{2}a_{8}^{c}\right)}\,,\\
\epsilon_{3/2}e^{i\hat{\phi}} & = & -\epsilon_{KM}\\
 &\times & \frac{ a_{1}+R_{\pi K}a_{2}+\frac{3}{2}\left(a_{10}^{u}+r_{\chi}^{K}a_{8}^{u}
 +R_{\pi K}\left(a_{9}-a_{7}\right)\right)}{\left(a_{4}^{c}-\frac{1}{2}a_{10}^{c}\right)+r_{\chi}^{K}\left(a_{6}^{c}-\frac{1}{2}a_{8}^{c}\right)}\,,\nonumber \\
\epsilon_{T}e^{i\hat{\phi}_{T}} & = & -\epsilon_{KM}\frac{a_{1}+\frac{3}{2}\left(a_{10}^{u}+r_{\chi}^{K}a_{8}^{u}\right)}{\left(a_{4}^{c}-\frac{1}{2}a_{10}^{c}\right)+r_{\chi}^{K}\left(a_{6}^{c}-\frac{1}{2}a_{8}^{c}\right)}\,,\label{eq:eT}\\
qe^{i\omega} & = & -\frac{3}{2\epsilon_{KM}}\\
 &\times & \frac{a_{10}^{c}+r_{\chi}^{K}a_{8}^{c}+R_{\pi K}\left(a_{9}-a_{7}\right)}{a_{1}+R_{\pi K}a_{2}+\frac{3}{2}\left(a_{10}^{u}+r_{\chi}^{K}a_{8}^{u}+R_{\pi K}\left(a_{9}-a_{7}\right)\right)}\,,\nonumber \\
q_{C}e^{i\omega_{C}} & = & -\frac{3}{2\epsilon_{KM}}\frac{a_{10}^{c}+r_{\chi}^{K}a_{8}^{c}}{a_{1}+\frac{3}{2}\left(a_{10}^{u}+r_{\chi}^{K}a_{8}^{u}\right)}\,,\label{eq:QCDF-q}
\end{eqnarray}
 where the notations
\begin{eqnarray*}
R_{\pi K} & = & \frac{A_{K\pi}}{A_{\pi K}},\quad\epsilon_{KM}=\frac{|\lambda_{u}|}{|\lambda_{c}|}\simeq\frac{\lambda^{2}}{2}\,,
\end{eqnarray*}
are defined.

The parameters would be process-dependent due to the scale $\mu$
of the $a_{i}$. The scale $\mu$ could be different from the assumed
$\mu=m_{b}$ and also depends on the process. These complicate relations
are represented by the $\mu_{i}$ with $i$ as an index for a specific process.
Therefore, these parameters are interpreted as functions
of $\mu_{i}$ and calculated through the above equations Eqs.(\ref{eq:QCDF-P}-\ref{eq:QCDF-q}).
In this work, we will assume that $P$, $\epsilon_{a}$,
$\epsilon_{3/2}$, $\epsilon_{T}$, $q$ and $q_{C}$, $\omega$, and $\omega_{C}$ are pure perturbative and
allow $\phi_{a}$, $\phi$, and $\phi_{T}$ to contain
both perturbative parts, $\hat{\phi}_{a}$, $\hat{\phi}$, and $\hat{\phi}_{T}$,
and nonperturbative parts, $\bar{\phi}_{a}$, $\bar{\phi}$, and $\bar{\phi}_{T}$:
\begin{eqnarray*}
\phi_{a} & = & \hat{\phi}_{a}+\bar{\phi}_{a}\,,\\
\phi & = & \hat{\phi}+\bar{\phi}\,,\\
\phi_{T} & = & \hat{\phi}_{T}+\bar{\phi}_{T}\,.
\end{eqnarray*}
This is because only $\phi_{a}$, $\phi$, and $\phi_{T}$ can directly
involve in the direct CP asymmetries.

\section{Analysis\label{sec:Analysis}}

In order to determine the eight parameters, $\gamma,$ $\bar{\phi}_{a}$,
$\bar{\phi}$, and $\bar{\phi}_{T}$, and $\mu_i$ with $i=1,\cdots,4$, introduced in the above, we
employ the following fitting procedure by using the eight measurements
of four branching ratios and four direct CP asymmetries. The input
data given in Table \ref{tab:Input-parameters} are used to calculate
the amplitude parameters given in Eqs.(\ref{eq:QCDF-P}-\ref{eq:QCDF-q}),
which are then substituted into the Eqs.(\ref{eq:Br1}-\ref{eq:ACP4}
). The data are separated into four sets, one set for one process.
Each set contains one branching ratio and one asymmetry: Set-1:$Br(\pi^{-}\bar{K}^{0})$
and $A_{CP}(\pi^{-}K_{s}^{0})$; Set-2:$Br(\pi^{0}K^{-})$ and $A_{CP}(\pi^{0}K^{-})$;
Set-3:$Br(\pi^{+}K^{-})$ and $A_{CP}(\pi^{+}K^{-})$; Set-4:$Br(\pi^{0}\bar{K}^{0})$
and $A_{CP}(\pi^{0}K^{0})$. Their experimental data are present in
Table \ref{tab:Predictions}, in which only the PDG2016 data \citep{Patrignani:2016xqp}
are used and the HFAG2016 data \citep{Amhis2017} are listed for reference.
The least squares fit method is used. The $\chi^{2}$ function is defined
as
\begin{align}
\chi_{i}^{2}(\vec{\theta}) & =\sum_{j=1}^{2}\frac{(\textit{\textit{O}}_{ij}^{exp}-\textit{O}_{ij}^{th}(\vec{\theta}))^{2}}{\sigma_{ij}^{2}}\,,\\
\chi_{tot}^{2}(\vec{\theta}) & =\sum_{i=1}^{4}\chi_{i}^{2}(\vec{\theta})\,,
\end{align}
where $\chi_{i}^{2}$ is the $\chi^{2}$ function for the $i$-th
set of data: $i=1$ for $B^{-}\to\pi^{-}\bar{K}^{0}$, $i=2$ for
$B^{-}\to\pi^{0}K^{-}$, $i=3$ for $\bar{B}^{0}\to\pi^{+}K^{-}$
and $i=4$ for $\bar{B}^{0}\to\pi^{0}\bar{K}^{0}$. The fitting parameters
are $\vec{\theta}=(\mu_{i},\,i=1,\cdots,4;\,\gamma,\,\bar{\phi}_{a},\,\bar{\phi},\,\bar{\phi}_{T})$.
The scale variable $\mu_{i}$ is defined for $i$-th data set.
The experimental measurements $O_{ij}^{exp}$, the theoretical predictions
$O_{ij}^{th}$, and the experimental errors $\sigma_{ij}$ are for
the $j$-th measurement of the $i$-th data set, $j=1$ for the branching
rate and $j=2$ for the asymmetry. For each fitting procedure, only
two measurements and one fitting parameter are used.
The degree of freedom is equal to one.
However, the nonlinear functional form of the expressions for the branching
ratios and the asymmetries makes the fitting to be a nonlinear fit problem.
It is not appropriate by using the standard least $\chi^{2}$ fit method,
in which all the parameters are simultaneously determined by calculating the best-fit value of the $\chi^{2}$ function.
To improve the task of fitting, we design the following iterative strategy similar
to the Marquardt method \citep{Press2010}.
The value of $\chi^{2}=\chi_{i}^{2},\chi_{tot}^{2}$ for each fitting procedure is very close to zero, $\chi^{2}\simeq0$.

\subsection{Fitting Strategy}

The following fitting strategy is employed to determine $\gamma$, $\bar{\phi}_{a}$, $\bar{\phi}$, $\bar{\phi}_{T}$ and $\mu_{i}$, $i=1,\cdots,4$.
\begin{enumerate}
\item Choose a reference value $\gamma_{init}$ for the weak angle $\gamma$.
Here, $\gamma_{init}={72.1}^{\circ}$ is chosen for the world averaged value
$(72.1_{-5.8}^{+5.4})^{\circ}$\citep{Amhis2017}.
\item Calculate the $\chi^{2}$ value for the data set $i=1,\cdots,4$ by varying
the factorization scale variable $\mu_{i}$ within the range $m_{b}/2\le\mu_{i}\le2m_{b}$
to find a least $\chi_{i,\mu}^{2}$ at the temporal $\mu_{i}^{T}$.
\item Calculate the $\chi^{2}$ value for the data set $i$ by varying one of
the strong phase variables $\bar{\phi}_{a},\bar{\phi},\bar{\phi}_{T}$
within the range $-180^{\circ}\le\bar{\phi}_{a},\bar{\phi},\bar{\phi}_{T}\le180^{\circ}$
with the previous temporal $\mu_{i}^{T}$ to find a least $\chi_{i,\phi}^{2}$
at the temporal $\bar{\phi}_{a,i}^{T},\bar{\phi}_{i}^{T},\bar{\phi}_{T,i}^{T}$.
\item Repeat procedures 3 and 4 until $\chi_{i,\mu}^{2}=\chi_{i,\phi}^{2}$
and obtain the fitted values of $\mu_{i},\bar{\phi}_{a,i},\bar{\phi}_{i},\bar{\phi}_{T,i}$.
\item The fitted values of $\mu_{i},\bar{\phi}_{a,i},\bar{\phi}_{i},\bar{\phi}_{T,i}$
are used to calculate the least $\chi^{2}$ for all data by varying
$-180^{\circ}\le\gamma\le180^{\circ}$ to find a least $\chi_{tot}^{2}$
at $\gamma_{fit}$.
\item Use $\gamma_{fit}$ to repeat procedures 2-6 until the $\chi_{tot}^{2}$
reaches a stable least value.
\item Use the criteria $\chi^{2}\le\sqrt{2\nu}$ to set the upper and
lower bounds of each parameter. $\nu$ is the degree of freedom of
the used data points and defined as $\nu=N-n$ with $N$ the number
of data point and $n$ the number of the fitting parameters. For the
procedures 3 and 4, $N=2$ and $n=1$ and $\nu=1$. The bounds of
$\gamma_{fit}$ are determined by using $\chi^{2}\le\sqrt{2\nu}$
for $\nu=N-n=4$ with $N=8$ and $n=4$.
\end{enumerate}

\subsection{Fitted Results}

The fitted results for $\mu_i$, $\bar{\phi}_{a},\bar{\phi},\bar{\phi}_{T}$
and $\gamma$ are listed in the Table\ref{tab:fit-data}.
The parameters calculated at $\mu_{i}$ are list in Table\ref{tab:amplitude parameters}.
For comparison, the parameters calculated for $m_{b}/2\le\mu\le2m_{b}$
with its central value at $\mu=\mu_{0}=m_{b}$ are also listed (denoted
as $\mu_{0}$) in the same table. The fitted scales $\mu_{i}$ are
used to calculate the averaged scale $\mu_{av}$. (See below explanation
for $\mu_{av}$.) The predictions of QCDF for branching ratios and
asymmetries are calculated at $m_{b}/2\le\mu\le2m_{b}$ and $\mu=\mu_{av}\pm\sigma_{av}$,
denoted as Naive and Improved columns in Table \ref{tab:Predictions}.
Using the fitted parameters to calculate the predictions for the branching
ratios and asymmetries are given in the Fit column in Table\ref{tab:Predictions},
where the first errors are from $\mu_{i}$ and the second errors from
the phase factors, $\bar{\phi}_{a},\bar{\phi},\bar{\phi}_{T}$.
For comparisons, the S4 column in Table\ref{tab:Predictions} are quoted from Ref.\citep{2003NuPhB.675..333B},
in which the predictions are calculated at the twist-3 two parton and NLO in $\alpha_s$ order
in the S4 scenario.

Comments of the above analysis are given as follows.
\begin{itemize}
\item In QCDF, the factorization scale $\mu$ separates the perturbative
physics from the nonperturbative. Priority, it is not possible to
guest what scale is appropriate for a process. In the $B$ decays,
$\mu=m_{b}$ is a convenient but not absolute choice. Any $\mu$ within
the range $m_{b}/2\le\mu\le2m_{b}$ is allowable. According to the
fit result, different processes require different factorization
scales.
\item The branching ratios are sensitive to the factorization scale $\mu$,
while the asymmetries are sensitive to the strong phases $\phi_{a}$,
$\phi$, and $\phi_{T}$ .
\item The predictions for branching ratios with $m_{b}/2\le\mu\le2m_{b}$
can cover the experimental data within errors.
\item The predictions for $A_{CP}(B^{-}\to\pi^{-}K_{s}^{0})$ and $A_{CP}(\bar{B}^{0}\to\pi^{+}K^{-})$
are in opposite sign to the data, and the predictions for $A_{CP}(B^{-}\to\pi^{0}K^{-})$
and $A_{CP}(\bar{B}^{0}\to\pi^{0}K^{0})$ are in the same sign to
the data.
\item The predictions for four asymmetries have consistent magnitudes with
the experimental data.
\item The perturbative parts of $\phi_{a}$, $\phi$, and $\phi_{T}$ are
in opposite sign to their nonperturbative parts.
\item The experimental data favor a large negative $\phi_{a}$ and moderate negative $\phi$ and moderate
positive $\phi_{T}$. Especially, $\bar{B}^{0}\to\pi^{0}K^{0}$ mode
favors $\phi\lesssim\phi_{T}$ to compensate $\epsilon_{3/2}\gtrsim\epsilon_{T}$.
\item The major uncertainties come from the $A_{CP}(B^{-}\to\pi^{-}K^{0})$ for $\bar{\phi}_{a}$
and $A_{CP}(\bar{B}^{0}\to\pi^{0}K^{0})$ for $\bar{\phi}$ and $\bar{\phi}_{T}$.
However, the more accurate asymmetries $A_{CP}(B^{-}\to\pi^{0}K^{-})$
and $A_{CP}(\bar{B}^{0}\to\pi^{+}K^{-})$ require $\bar{\phi}_{a}$,
$\bar{\phi}$, and $\bar{\phi}_{T}$ being process-dependent.
\item $\bar{\phi}_{a}$ is determined by data set 1 and needs specific values
for the data sets 2, 3, 4, respectively.
\item $\bar{\phi}$ and $\bar{\phi}_{T}$ are equal and determined
by data set 4 and they need specific values for the data sets 2 and
3.
\item The better parametrization would
be $\Phi_{a}=\bar{\phi}_{a}-\gamma$, $\Phi=\bar{\phi}-\gamma$, and
$\Phi_{T}=\bar{\phi}_{T}-\gamma$. This allows for further exploration
of the NP effects.
\item Only uncertainties
from the scale variables are taken as theoretical
errors. The uncertainties from the input
data are completely neglected.
\end{itemize}

\section{Discussions\label{sec:Discussions}}

To examine whether the puzzle could be resolved in our approach, we
choose the following topics to discuss.

\subsection{Basic Results}

\paragraph{Fitting }

The fit of each data set gives $\chi^{2}\simeq0$
for one degree of freedom ($\nu=1$). According to the least squares
method (see e.g. \citep{doi:10.1137/0111030}), a very small $\chi^{2}\ll1$
could mean either (i) our model is valid, or (ii) the experimental
errors are too large, or (iii) the data is too good to be true. Since
a poor model can only increase $\chi^{2}$, a too-small value of $\chi^{2}$
cannot be indicative of a poor model. This implies that our model
could be a good model for the data. The fitted results show that (i)
the factorization scale variable $\mu$ is process-dependent, (ii)
the strong phase variables $\phi_{a}$, $\phi$, $\phi_{T}$ are process-dependent.
As a result, the eleven parameters are process-dependent as shown in Table \ref{tab:amplitude parameters}.
This observation is in contradiction to the usual assumption that these
parameters are process-independent. If this founding that
the parameters are process-dependent is true, then the
puzzle found by assuming the parameters to be process-independent
would be questionable.

\paragraph{Averaged Analysis}

From Table \ref{tab:fit-data}, the factorization scales $\mu_{i}$
of four processes are averaged as $\mu_{av}=4.96\pm1.44$GeV, where
the averaged factorization scale $\mu_{av}\pm\sigma_{av}$ is calculated
by means of the weighted mean method
\begin{eqnarray}
\mu_{av} & = & \frac{\sum_{i}\sigma_{i}\mu_{i}}{\sum_{i}\sigma_{i}}\,,\;\sigma_{av}=\left(\frac{1}{\sum_{i}\sigma_{i}^{2}}\right)^{1/2}\,.
\end{eqnarray}
The $\mu_{av}$ denotes the specific energy scale relevant to the
$B\to\pi K$ decays. For distinguishing, we call the parameters calculated
by $\mu_{0}=m_{b}$ as the ``Naive'' predictions and those calculated
at $\mu_{av}$ as the ``Improved'' predictions. The uncertainties
of $\mu_{av}$ come from the experimental data and those of $\mu_{0}$
from the assumption of QCDF. The $\mu_{av}$ is more appropriate
than by comparing their predictions with the corresponding data as given in Table \ref{tab:Predictions}.
The parameters calculated at $\mu_{av}\pm\sigma_{av}$ are present in the column $\mu_{av}$ of Table \ref{tab:amplitude parameters}.

\paragraph{Three Parton Effects}

For comparison, the amplitude parameters with twist-3 three parton
NLO corrections calculated in this work and similar terms with the
twist-3 two parton NLO corrections calculated in Ref. \citep{BENEKE2001245}
are present in the second column and the third column of Table \ref{tab:amplitude parameters}.
Only $P^{\prime}$, $\epsilon_{T}$, $q_{C}$, and $\omega_{C}$ have
significant differences between twist-3 two parton and twist-3 three
parton calculations. The three parton penguin term $P^{(3)}=51.4$eV
has about $15.5\%$ enhancement than the two parton $P^{(2)}=44.5$eV.
The most significant difference comes from the $\epsilon_{T}$. The
three parton $\epsilon_{T}^{(3)}=14.6(\%)$ is about two third of
the two parton $\epsilon_{T}^{(2)}=22.0(\%)$. This can be easily
understood due to the enhanced effects from the three parton corrections,
the denominator of Eq.(\ref{eq:eT}) contains $r_{\chi}(a_{6}-\frac{1}{2}a_{8})$.
The ratio $r=(\epsilon_{3/2}-\epsilon_{T})/\epsilon_{T}\sim C^{\prime}/T^{\prime}$
has been suggested to be an important index for distinguishing whether
SM can or can not explain the puzzle. Let's compare the theoretically
predicted values: the naive $\epsilon_{3/2}(\%)=(22.7\pm7.0)_{th}$
and $\epsilon_{T}(\%)=(14.6\pm4.0)_{th}$, the improved $\epsilon_{3/2}(\%)=(24.2\pm3.0)_{th}$
and $\epsilon_{T}(\%)=(15.4\pm1.8)_{th}$, and the fit values,
$\epsilon_{3/2}(\%)=(24.3\pm1.1)_{fit}$ and $\epsilon_{T}(\%)=(15.4\pm0.6)_{fit}$,
which result in
\begin{eqnarray*}
r_{fit} & \simeq & (0.59\pm0.08)_{fit}\\
r_{th} & = & (0.55\pm0.57)_{th}(Naive)\\
r_{th}^{\prime} & = & (0.57\pm0.24)_{th}(Improved)
\end{eqnarray*}
The Naive $r_{th}$, the improved $r_{th}^{\prime}$,
and the fit $r_{fit}$ imply that SM can explain the puzzle at $1\sigma$, $2\sigma$
and $7.4\sigma$, respectively. For reference, the two parton prediction
gives $r^{(2)}\simeq0.2$ as calculated from the second column of
Table \ref{tab:amplitude parameters}, the values quoted from Ref.
\citep{BENEKE2001245} .

As for $q$, the three parton $q^{(3)}=0.33$ is about four times
the two parton $q^{(2)}=0.083$. The three parton predictions can
accommodate the data without additional EW penguin contributions from
other sources as the two parton predictions required. On the other
hand, the three parton $\omega_{C}^{(3)}=-7.9^{\circ}$ is about one
sixth of the two parton $\omega_{C}^{(2)}=-54^{\circ}$.

The enhancements of the predictions for the branching ratios by the three parton corrections could be seen by
comparing the Improved and the S4 columns in Table \ref{tab:Predictions}.
The S4 predictions quoted from \citep{BENEKE2001245} are calculated at the twist-3 two parton order in the S4 scenario.

\subsection{Predictions}

\paragraph{Power Counting Puzzle}

In literature, the simple version of the $B\to\pi K$ puzzle is based
on the analysis for the predictions of the SM by means of a power
counting of the amplitude parameters \citep{Beaudry2018,2007IJMPA..22.2057I,1995PhRvD..52.6356G,1995PhRvD..52.6374G}.
Under the standard parametrization of the CKM matrix elements, we
have $\lambda_{u}\sim O(\lambda^{4})$ and $\lambda_{c}\sim O(\lambda^{2})$,
where $\lambda=0.22$ is the sine of the Cabibbo angle. If the $SU(3)$
flavor symmetry is valid, then we have $\epsilon_{a}\sim\epsilon_{KM}\sim O(\lambda^{2})$,
$\epsilon_{3/2}\sim\epsilon_{T}\sim O(\lambda)$, $q\sim q_{C}\sim1$,
$\epsilon_{3/2}-\epsilon_{T}\sim O(\lambda^{2})$ \citep{Beaudry2018,2007IJMPA..22.2057I,1995PhRvD..52.6356G,1995PhRvD..52.6374G}.
Combining these facts, we have the following Standard Model (SM) predictions
$A_{CP}(B^{-}\to\pi^{0}K^{-})=A_{CP}(\bar{B}^{0}\to\pi^{+}K^{-})\sim O(\lambda)$
and a vanishing $\Delta_{th}A_{CP}=0$ up-to $O(\lambda^{2})$, if
the strong phases are of similar orders, $\phi\sim\phi_{T}$, $\omega\sim\omega_{C}$
\citep{Buras:2003dj}.

Because the $\Delta A_{CP}$ is about $O(\lambda)$, at which SM predicts
a vanishing $\Delta A_{CP}$ with errors $O(\lambda^{2})$. However,
the direct CP asymmetries of four processes are all of $O(\lambda^{2})$.
Therefore, in order to reveal the mystery of the puzzle, it needs
the calculation accuracy up-to $O(\lambda^{2})$

By order of magnitudes, the twist-3 power corrections are of $O(\Lambda/m_{B})\sim O(\lambda^{2})$
and the NLO $\alpha_{s}$ corrections are of $O(\alpha_{s}^{2}(m_{b}))\sim O(\lambda^{2})$.
This implies that QCDF with complete twist-3 power corrections and
NLO corrections has the precision of $O(\lambda^{2})$. That ensures
that our calculations are able to distinguish the different terms
of the asymmetries. As a result, we may be able to figure out the
crucial differences which lead to the puzzle.

To make the above explanation clear, let's separately calculate the
four terms in the square bracket of each direct CP asymmetry in Eqs.
(\ref{eq:ACP1}-\ref{eq:ACP4}). The results are present in Table
\ref{tab:Direct-CP-asymmetry}, the columns $1,\cdots,4$, denote
the j-th term of each $A_{CP}$ equation. For $A_{CP}(\pi^{0}K^{0})$,
the terms are separated by different lines. The fourth term contains the last four terms. Now, we may understand
why the puzzle may not exist:
\begin{enumerate}
\item Although $\epsilon_{a}$ is much smaller than $P^{\prime}$, but it
should not be neglected for $A_{CP}$.
\item Most of the first and second terms are of $O(\lambda^{2})$ and the
third and fourth terms of $O(\lambda^{3})$. The largest term is $-\epsilon_{3/2}\sin(\phi)\sin(\gamma)$
of $A_{CP}(\pi^{0}K^{-})$. The second large term is $-\epsilon_{T}\sin(\phi_{T})\sin(\gamma)$
of $A_{CP}(\pi^{+}K^{-})$. Both terms have different signs and magnitudes.
\item The puzzle assumes $N_{2}=N_{3}=2$, but, in fact, $N_{2}=1.65$ and
$N_{3}=1.94$.
\item The puzzle requires $|\epsilon_{3/2}-\epsilon_{T}|\sim O(\lambda^2)$.
On the other hand, the data favor $|\epsilon_{3/2}-\epsilon_{T}|\sim O(\lambda)$.
\end{enumerate}
In summary, the power counting can only differentiate $O(\lambda)$ and can not correctly distinguish the signs of the parameters.
The $SU(3)_{f}$ needs to be broken to explain the $\Delta A_{CP}\neq0$,
which comes from the largest terms of $A_{CP}(\pi^{0}K^{-})$ and
$A_{CP}(\pi^{+}K^{-})$ being unequal and in opposite signs.

\paragraph{Mixing Induced CP Asymmetry}

We now test MQCDF model by the mixing induced
CP asymmetry $S_{\pi^{0}K_{S}}=0.58\pm0.06$ \citep{Patrignani:2016xqp}.
Under the invariance of CPT, the time dependent CP asymmetry $A_{CP}(t)$
for $\bar{B}^{0}\to\pi^{0}K^{0}$ is given by \citep{Patrignani:2016xqp}
\begin{eqnarray}
A_{CP}(t) & = & \frac{\Gamma_{\bar{B}\to f}(t)-\Gamma_{B^{0}\to f}(t)}{\Gamma_{\bar{B}\to f}(t)+\Gamma_{B^{0}\to f}(t)}\\
 & = & \frac{-C_{f}\cos(\Delta m_{d}t)+S_{f}\sin(\Delta m_{d}t)}{\cosh(\frac{\Delta\Gamma_{d}}{2}t)+A_{f}^{\Delta\Gamma}\sinh(\frac{\Delta\Gamma_{d}}{2}t)}\,,\nonumber
\end{eqnarray}
where the $B^{0}-\bar{B}^{0}$ system has mass difference $\Delta m_{d}$
and width difference $\Delta\Gamma_{d}$ for the mass eigenstate $|B^{0}\rangle$
and $|\bar{B}^{0}\rangle$. The quantities are given by
\begin{eqnarray}
C_{f} & = &\frac{1-|\lambda_{f}|^{2}}{1+|\lambda_{f}|^{2}}\,,\\
S_{f} & = &\frac{2\text{Im}(\lambda_{f})}{1+|\lambda_{f}|^{2}}\,,\\
A_{f}^{\Delta\Gamma} & = &-\frac{2\text{Re}(\lambda_{f})}{1+|\lambda_{f}|^{2}}\,,
\end{eqnarray}
where $\lambda_{f}$ is defined as
\begin{widetext}
\begin{eqnarray}
\lambda_{f}  =  -e^{-2i\beta}\left[\frac{(1+\epsilon_{a}e^{i\phi_{a}}e^{-i\gamma}+(\epsilon_{3/2}e^{i\phi}-\epsilon_{T}e^{i\phi_{T}})e^{-i\gamma}+(\epsilon_{T}q_{C}e^{i(\phi_{T}+\omega_{C})}-\epsilon_{3/2}qe^{i(\phi+\omega)}))}{(1+\epsilon_{a}e^{i\phi_{a}}e^{i\gamma}+(\epsilon_{3/2}e^{i\phi}-\epsilon_{T}e^{i\phi_{T}})e^{i\gamma}+(\epsilon_{T}q_{C}e^{i(\phi_{T}+\omega_{C})}-\epsilon_{3/2}qe^{i(\phi+\omega)}))}\right]\,.
\end{eqnarray}
\end{widetext}
The bets-fit results are as follows:
\begin{enumerate}
\item By using the world averaged $\beta=(22.5\pm4.4\pm1.2\pm0.6)^{\circ}$,
the calculations give $S_{f}=0.57\pm0.13$ and $C_{f}=-0.01$ and
$A_{f}^{\Delta\Gamma}=0.82\pm0.09$.
\item By using the above expression and the fitted parameters for $\pi^{0}K_{S}$
mode given in the column $\mu_{4}$ of Table\ref{tab:amplitude parameters},
the fit to $S_{f}=0.58\pm0.06$ gives $\beta=(22.92_{-2.13}^{+2.08})^{\circ}$,
which is in agreement with the world averaged $\beta=(22.5\pm4.4\pm1.2\pm0.6)^{\circ}$.
\end{enumerate}

It implies that $S_{\pi^{0}K_{S}}=0.58\pm0.06$ and $\beta=(22.5\pm4.4\pm1.2\pm0.6)^{\circ}$ are consistent according to
the MQCDF model.

\paragraph{Ratios of Modes}

Three ratios, $R$, $R_{c}$ and $R_{n}$ have been widely
used for demonstration of the $B\to\pi K$ puzzle. We now compare
their theoretical ($R^{th}$ in the Improved column) and experimental
values ($R^{exp}$ in the 2016 column)in Table \ref{tab:Ratios},
which are calculated according to the Improved column in Table \ref{tab:Predictions}
and the experimental data, respectively.

From the above calculations, we may notice the following interesting
points:
\begin{enumerate}
\item For the ratio $R$, the central value of the predicted value $R^{th}=0.91$
is vary close to that of the experimental value $R^{exp}=0.89$ within
$3\%$.
\item The ratio of central values show $R_{n}^{th}/R_{c}^{th}\simeq1+0.9\%$
and $R_{n}^{exp}/R_{c}^{exp}=1+9.2\%$. Both are compatible within
$10\%$ despite of their respective uncertainties. The discussion
about the $10\%$ difference is left to the Sec.\ref{sec:Conclusion}.
\end{enumerate}
In summary, the comparisons between the theoretical predictions of
this work and the experimental data for these three ratios
show no similar large discrepancies as claimed in \citep{Buras:2003dj,Buras:2003yc,Buras:2004ub}.

\subsection{Comparisons With Other Approaches}

\paragraph{Flavor Symmetry Approach}

Many studies have employed the flavor symmetry approach \citep{Zeppenfeld:1980ex,1995PhLB..341..379B,1995PhLB..360..138B,1995PhRvD..52.6356G,1998PhRvD..57.2752F,Gronau:1998fn,1126-6708-1999-02-014,Buras:2003dj,2003PhLB..572...43G,Buras:2003yc,Buras:2004ub,Chiang:2004nm,Fleischer:2008wb},
which is an extension of the isospin symmetry approach. The amplitudes
are expressed in terms of parameters which satisfy the symmetry. To
conserve the symmetry, the parameters would be process-independent.
By means of many sophisticated arguments, many important insights
for the flavor physics have been derived. The arguments heavily rely
on the assumption that the common parameters involved in different processes
are universal such that they can be used for predictions. However,
these arguments would be questionable if the parameters are process-dependent
according to our founding in this work.

\paragraph{Global-Fit Approach}

In literature, there are studies \citep{Baek:2007yy,Baek:2009pa,Beaudry2018}
by employing the global-fit for analysis of the $B\to\pi K$ puzzle. The
basic assumption of this fitting approach is that there would exist
some symmetries such that the amplitudes can
be parameterized in terms of universal parameters. However, this
method has its intrinsic uncertainties. Since there may exist many (perhaps
infinite) possible best-fit results, it is difficult to distinguish
which result is the correct one.
In addition, it is difficult to explain the underlying physics for the best-fit results.
In our approach, there is no such problem.

\paragraph{NLO PQCD Approach}

Similar studies have been made by employing the PQCD approach \citep{Li:2005kt,2014ChPhC..38c3101B,2011PhRvD..83c4023L}.
The complete NLO calculations \citep{2014ChPhC..38c3101B} are improved
than the partial NLO ones \citep{Li:2005kt}. The complete NLO predictions
\citep{2014ChPhC..38c3101B} for the branching ratios and direct CP
asymmetries are compatible with the experimental data. The authors
of Ref. \citep{2011PhRvD..83c4023L} indicated that some nonperturbative
strong phase from the Glauber gluons are necessary. Although the QCDF
and the PQCD approaches are based on different factorization assumptions,
their calculations up-to NLO and twist-3 order agree within theoretical
uncertainties. This can be seen by comparing the Improved and PQCD columns in Table\ref{tab:Predictions}.

\paragraph{Final State Interactions}

The final state interactions of the $B\to\pi K$ decays are introduced
to account for strong phases (see e.g. \citep{2005PhRvD..71a4030C}).
This approach uses the QCDF predictions as the reference by adding
the final state interaction effects. The final state interactions
occur through long distance inferences between different final states.
The calculations introduce nonperturbative parameters which are determined
by a best-fit to the data. Since the QCDF calculations contain final
state interactions in the $a_{i}$ functions, there would exist double
counting effects in this approach \citep{2003NuPhB.675..333B}. On
the other hand, our model has avoided this problem.

\paragraph{Endpoint Divergences}

There exist endpoint divergent terms in the standard QCDF calculations.
To regularize the divergences, the following model is introduced \citep{2003NuPhB.675..333B}
\[
X_{A}=(1+\rho_{A}e^{i\phi_{A}})\ln(\frac{m_{B}}{\Lambda_{h}});\quad\rho_{A}\le1,\quad\Lambda_{h}=0.5\text{GeV}.
\]
The parameters $\rho_{A}$ and $\phi_{A}$ are determined by a global-fit
to the data. Because the factorization formula Eq.(\ref{eq:fact-formula})
is free from these end-point divergences \citep{Yeh:2007fi}, no such
terms need to be considered in our calculations. Since the phase $\phi_{A}$
is associated with the annihilation terms, the explanations for the
puzzle are different from ours.

\subsection{Isospin Symmetry Breaking }

\paragraph{Broken Isospin Symmetry}

The isospin symmetry is conserved under the weak interactions.
If the isospin symmetry is also conserved by the QCD,
then the $B\to\pi K$ amplitudes would obey the quadrangle relation Eq.(\ref{eq:quadrangle relation})
\citep{Nir:1991cu,1995PhRvD..52.6374G}.
This relation leads to the ratio $R_{c}/R_{n}=1$, which is shown
by the central value of the QCDF prediction (the Improved) $R_{c}/R_{n}\simeq1$ in Table \ref{tab:Ratios}.
However, the central values of the last experimental data show that
$R_{c}/R_{n}=1.1>1$. How does this imply for the quadrangle relation?
To answer this, we may calculate the following ratio for the amplitudes
listed in Table \ref{tab:Amplitudes}
\begin{eqnarray}
r & = & \frac{A(\pi^{-}K^{0})+\sqrt{2}A(\pi^{0}K^{-})}{A(\pi^{-}K^{+})+\sqrt{2}A(\pi^{0}K^{0})}=\frac{A_{3/2}(B^{-})}{A_{3/2}(\bar{B}^{0})}\,,\\
r_{exp} & = & (0.87\pm0.15)e^{-i23^{\circ}}\,,\\
r_{th} & = & (1.04\pm1.0)e^{i0.4^{\circ}}\,,
\end{eqnarray}
where $r_{exp}$ is calculated by the Fit column in Table \ref{tab:Amplitudes}
and the $r_{th}$ is calculated by the Improved column in the same
table. The experimental data imply that the amplitudes would not obey
the quadrangle relation. It is in contradiction to the theoretical
assumption, $r_{exp}\neq r_{th}$. Within uncertainties, we obtain
$r_{exp}\neq1$ at $6\sigma$ significance. That is $A_{3/2}(B^{-})\neq A_{3/2}(\bar{B}^{0})$.
This shows that isospin symmetry needs to be broken for explaining the data.
Otherwise, the puzzle would remain.
In other words, the $B\to\pi K$ puzzle is solved by the broken isospin symmetry.

The isospin symmetry is broken by (i) the process-dependent factorization
scale $\mu_{i}$, and (ii) the process-dependent non-vanishing nonperturbative
phases $\bar{\phi}_{a}$, $\bar{\phi}$, $\bar{\phi}_{T}$.

It is interesting to note that the mass differences $\Delta m_{q}=|m_{d}-m_{u}|\simeq2.5$ MeV
with $m_{u(d)}$ the up (down) quark's mass, $\Delta m_{\pi}=|m_{\pi^{0}}-m_{\pi^{\pm}}|\simeq5$ MeV
with $m_{\pi^{0}(\pi^{\pm})}$ the pion's mass, $\Delta m_{K}=|m_{K^{0}}-m_{K^{\pm}}|\simeq4$ MeV
with $m_{K^{0}(K^{\pm})}$ the kaon's mass, or $\Delta m_{B}=|m_{B^{0}}-m_{B^{\pm}}|\simeq0.3$ MeV
with $m_{B^{0}(B^{\pm})}$ the $B^{0(\pm)}$ meson's mass can also
break the isospin symmetry. The largest effect comes from the mass
difference,$\Delta m_{\pi}\simeq5$ MeV or $\Delta m_{K}\simeq4$ MeV,
which contribute about $5\times10^{-5}$ of the ratio $r$.

To distinguish, we call the former as the ``dynamic'' and the latter as the ``static''  isospin symmetry
breaking.

\paragraph{Nonperturbative Strong Phases}

The three nonperturbative strong phases $\bar{\phi}_{a}$, $\bar{\phi}$,
$\bar{\phi}_{T}$ have a specific pattern in $B\to\pi K$ system as
shown in Table\ref{tab:fit-data}. The full strong phases, $\phi_{a}=\hat{\phi}_{a}+\bar{\phi}_{a}$,
$\phi=\hat{\phi}+\bar{\phi}$, $\phi_{T}=\hat{\phi}_{T}+\bar{\phi}_{T}$,
obey the pattern $|\phi_{a}|>|\phi|\simeq|\phi_{T}|$.
A large negative $\phi_{a}$ is observed. Since these three phases
are closely related to the weak angle $\gamma$, three phases could
be redefined as $\Phi_{a}=\phi_{a}-\gamma$, $\Phi=\phi-\gamma$,
$\Phi_{T}=\phi_{T}-\gamma$. From this respect, there exist three
possibilities
\begin{enumerate}
\item The puzzle is solved by SM, if the phase factors $\phi_{a}$, $\phi$,
$\phi_{T}$ are completely of strong interactions.
\item The puzzle is solved by NP, if the phase factors $\Phi_{a}$, $\Phi$,
$\Phi_{T}$ completely come from NP effects.
\item Both of 1 and 2 are possible.
\end{enumerate}
To clarify which scenario is correct relies on future studies.

\section{Conclusion\label{sec:Conclusion}}

In this work, we have developed an effective method for analyzing
the $B\to\pi K$ system by using the MQCDF model. The crucial roles
are played by the phase factors $\bar{\phi}_{a},\,\bar{\phi},\,\bar{\phi}_{T}$.
By the fitting strategy, we may determine their values definitely.
It was found that their values depend on the decay modes. It is possible
to extract a universal $\gamma$ from the data, whose value is in
good agreement with the world averaged value. The fit result was used
to extract the weak angle $\beta$ from the mixing induced CP asymmetry
$S_{\pi^{0}K_{S}^{0}}$. The extracted $\beta$ is in a good agreement
with the world averaged value. From these evidences, our proposed
model could completely solve the original puzzle defined in Sec.\ref{sec:Intro}.

The model could reconstruct the experimental
data at the amplitude level as shown in Table\ref{tab:Amplitudes}.
The isospin symmetry of the $B\to\pi K$ system would be broken dynamically by strong interactions,
such that the $B\to\pi K$ puzzle is solved.
The application of our method to other decay processes is straightforward.

\bibliographystyle{apsrev}
\phantomsection\addcontentsline{toc}{section}{\refname}\bibliography{BpiKPuzzle0827}
\newpage

\begin{figure*}
\caption{Three parton one loop diagrams for the vertex corrections of $T^{I}$
are shown. The red gluon line means the gluonic parton $g$ of the
three parton $|q\bar{q}g\rangle$ state of the external $M_{2}$ meson
state $|M_{2}\rangle$. The blue gluon line means the radiative loop
gluons. The black box means the effective four quark operator. The
other similar diagrams with the gluonic parton line connected to other
external or internal parton lines to the vertex or penguin radiative
loop corrections may also contribute. Similar three parton one loop
diagrams for the $T^{II}$ are not shown here. The figure refers to \citep{Yeh:2007fi}.}
\includegraphics[scale=0.8]{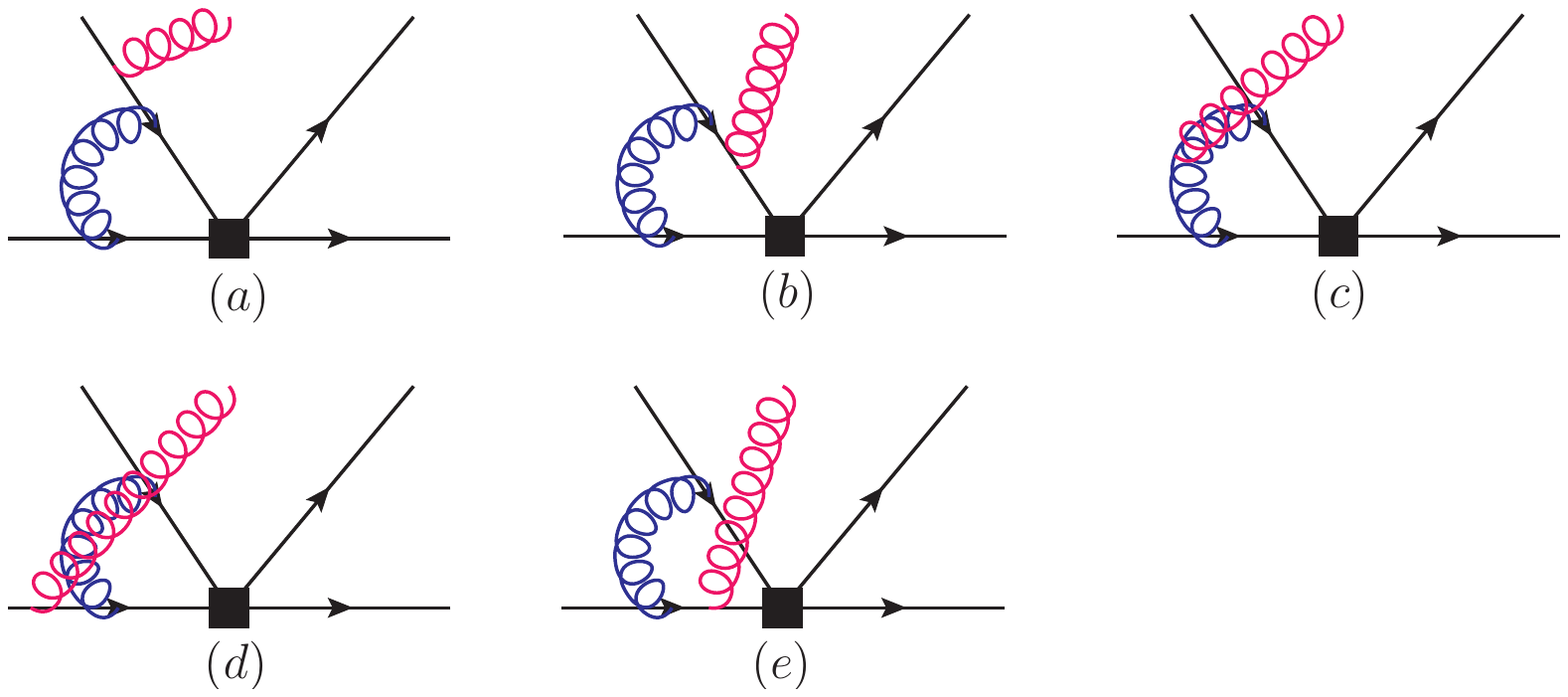}

\end{figure*}
\begin{table*}
\caption{\label{tab:Input-parameters}Input data \citep{Patrignani:2016xqp} for numerical calculations. The errors are neglected. The reason refers to the text.}
\begin{tabular*}{10cm}{c}

\begin{tabular}{p{2.5cm}|p{2.5cm}|p{2.5cm}|p{2.5cm}}
\hline
\multicolumn{4}{c}{B meson parameters}\tabularnewline
\hline
$\tau(B_{d}^{0})$(ps) &$\tau(B^{-})$(ps) & $m_{B^{\pm}}$(GeV) & $m_{B^{0}}$(GeV)\tabularnewline
\hline
$1.52$ & $1.64$ &$5.28$ & $5.28$\tabularnewline
\hline
\multicolumn{4}{c}{perturbative parameter}\tabularnewline
\hline
$\Lambda_{\text{\ensuremath{\bar{MS}}}}^{(5)}$(GeV) & $m_{b}$(GeV) & $m_{c}$(GeV) &$m_{s}$(GeV)\tabularnewline
\hline
0.225 & 4.2 & 1.3 & 0.09\tabularnewline
\hline
\end{tabular}

\\

\begin{tabular}{p{2.0cm}|p{2.0cm}|p{2.0cm}|p{2.0cm}|p{2.0cm}}
\multicolumn{5}{c}{decay constants and form factors}\tabularnewline
\hline
$f_{K}$(GeV) & $f_{\pi}$(GeV) & $f_{B}$(GeV) & $F_{0}^{B\pi}$(0) & $F_{0}^{BK}(0)$\tabularnewline
\hline
$0.16$ & $0.13$ & $0.2$ & $0.28$ & $0.34$\tabularnewline
\hline
\multicolumn{5}{c}{CKM matrix elements}\tabularnewline
\hline
$|V_{ub}|$ & $|V_{cb}$| & $|V_{us}|$ & $|V_{cd}|$ & $|V_{cs}|$ \tabularnewline
\hline
$0.0037$ & $0.042$ & $0.23$ & $0.22$ & $0.97$  \tabularnewline
\hline
\end{tabular}
\end{tabular*}
\end{table*}

\begin{table*}
\caption{\label{tab:fit-data}Four factorization
scales $\mu_i$ and the nonperturbative parts of the three strong phase parameters,
$\bar{\phi}_{a}$, $\bar{\phi}$, and $\bar{\phi}_{T}$ are determined
by a least squares fit method. The fitting strategy is described
in the text. The upper and lower bounds of the parameters are determined
by $\chi^{2}\le\sqrt{\nu}$ with $\nu$ the number of degrees of freedom.
The fit determines the weak angle $\gamma=(72.1_{-5.8}^{+5.7})^{\circ}$
.}
\begin{tabular}{p{2.0cm}|p{2.0cm}|p{2.0cm}|p{2.0cm}|p{2.0cm}}
\hline
decay modes & $\mu$(GeV) & $\bar{\phi}_{a}$(deg) & $\bar{\phi}$(deg) & $\bar{\phi}_{T}$(deg)\tabularnewline
\hline
\hline
$B^{-}\to\pi^{-}\bar{K}^{0}$ & $4.55_{-0.25}^{+0.28}$ & $-50_{-56}^{+32}$ & NA & NA\tabularnewline
\hline
$B^{-}\to\pi^{0}K^{-}$ & $5.25_{-0.37}^{+0.45}$ & $-50$ & $3.1_{-3.7}^{+3.6}$ & NA\tabularnewline
\hline
$\bar{B}^{0}\to\pi^{+}K^{-}$ & $5.6_{-0.26}^{+0.28}$ & $-115.1$ & NA & $16.3\pm1.4$\tabularnewline
\hline
$\bar{B}^{0}\to\pi^{0}K^{0}$ & $4.40_{-0.33}^{+0.35}$ & $-92$ & $23_{-40.6}^{+29.9}$ & $23_{-40.6}^{+29.9}$\tabularnewline
\hline
\end{tabular}
\end{table*}
\begin{table*}
\caption{\label{tab:amplitude parameters} Hadronic parameters are calculated at $\mu_{0}=4.2_{-2.1}^{+4.2}$, $\mu_{1}=(4.6\pm0.3)$,
$\mu_{2}=(5.3_{-0.4}^{+0.5})$, $\mu_{3}=(5.6\pm0.3)$, $\mu_{4}=(4.4_{-0.3}^{+0.4})$,
$\mu_{av}=(5.0\pm1.4)$. The unit is in GeV. For simplicity, only
central values are given for $\mu_{i}$, $i=1,\cdots,4$. The similar
calculations up-to twist-3 two parton NLO quoted from Ref.\citep{BENEKE2001245}
are listed for comparisons. The $\dagger$ term is calculated by using
the values given in Ref.\citep{BENEKE2001245}.}

\begin{tabular}{p{2.0cm}|p{2.0cm}|p{2.0cm}|p{2.0cm}|p{2.0cm}|p{2.0cm}|p{2.0cm}|p{2.0cm}}
\hline
parameters & $\mu_{0}$ & Ref\citep{BENEKE2001245} & $\mu_{1}$ & $\mu_{2}$ & $\mu_{3}$ & $\mu_{4}$ & $\mu_{av}$\tabularnewline
\hline
\hline
$|P^{\prime}|(eV)$ & $51.4_{-10.3}^{+16.1}$ & $44.5^{\dagger}$ & $50.0$ & 47.7 & 46.7 & 50.6 & $48.8_{-3.7}^{+6.0}$\tabularnewline
\hline
$\epsilon_{a}(\%)$ & $1.9\pm0.1$ & $1.9\pm0.1$ & $1.9$ & 1.9 & 1.9 & 1.9 & $1.9\pm0.1$\tabularnewline
\hline
$\epsilon_{3/2}(\%)$ & $22.7_{-6.4}^{+7.0}$ & $25.7\pm4.8$ & $23.5$ & 24.9 & 25.6 & 23.2 & $24.2_{-3.1}^{+2.5}$\tabularnewline
\hline
$\epsilon_{T}(\%)$ & $14.6_{-3.6}^{+4.0}$ & $22.0\pm3.6$ & $15.0$ & 15.8 & 16.0 & 14.8 & $15.4_{-1.8}^{+1.4}$\tabularnewline
\hline
$q(\%)$ & $60.1_{-4.4}^{+0.4}$ & $58.8\pm6.7$ & $60.2$ & 60.4 & 60.4 & 60.2 & $60.5_{-0.8}^{+0.2}$\tabularnewline
\hline
$q_{C}(\%)$ & $31.6_{-10.2}^{+3.2}$ & $8.3\pm4.9$ & $32.2$ & 33.1 & 33.5 & 32.0 & $33.0_{-3.0}^{+1.6}$\tabularnewline
\hline
$\hat{\phi}_{a}$(deg) & $21.5_{-2.5}^{+1.2}$ & $16.6\pm5.2$ & $21.3$ & 20.8 & 20.6 & 21.4 & $20.9_{-1.0}^{+1.1}$\tabularnewline
\hline
$\hat{\phi}$(deg) & $-10.2_{-1.7}^{+2.3}$ & $-10.2\pm4.1$ & $-10.4$ & -10.8 & -11.0 & -10.3 & $-10.6_{-0.6}^{+0.9}$\tabularnewline
\hline
$\hat{\phi}_{T}$(deg) & $-6.8_{-2.6}^{+3.8}$ & $-6.2\pm4.6$ & $-7.2$ & -7.8 & -8.0 & -7.0 & $-7.5_{-1.0}^{+1.5}$\tabularnewline
\hline
$\omega$(deg) & $0.4_{-0.7}^{+0.4}$ & $-2.5\pm2.8$ & $0.5$ & 0.6 & 0.6 & 0.5 & $0.5\pm0.2$\tabularnewline
\hline
$\omega_{C}$(deg) & $-9.3_{-12.8}^{+4.4}$ & $-54.2\pm44.2$ & $-8.6$ & -7.5 & -7.1 & -8.0 & $-7.9_{-3.2}^{+1.7}$\tabularnewline
\hline
\end{tabular}
\end{table*}
\begin{table*}
\caption{\label{tab:Predictions}Predictions and experimental data for branching ratios ($10^{-6}$) and direct CP asymmetries ($10^{-2}$). }
\begin{tabular}{p{3.0cm}|p{2.0cm}|p{2.0cm}|p{2.0cm}|p{2.0cm}|p{2.0cm}|p{2.0cm}|p{2.0cm}}
\hline
Decay modes & Naive & Improved & S4\citep{2003NuPhB.675..333B} & pQCD\citep{2014ChPhC..38c3101B} & Fit & PDG2016 & HFAG2016\tabularnewline
\hline
\hline
$Br(B^{-}\to\pi^{-}\bar{K}^{0})$ & $25.1_{-9.0}^{+18.2}$ & $22.6_{-3.3}^{+5.9}$ & 20.3 & $21.5_{-6.1}^{+8.0}$ & $23.9_{-0.9-0.2}^{+0.8+0.0}$ & $23.7\pm0.8$ & $23.79\pm0.75$\tabularnewline
\hline
$A_{CP}(B^{-}\to\pi^{-}K_{s}^{0})$ & $1.3\pm0.1$ & $1.3\pm0.0$ & 0.3 & $0.38_{-0.140}^{+0.097}$ & $-1.7_{-0.0-1.9}^{+0.0+1.7}$ & $-1.7\pm1.6$ & $-1.7\pm1.6$\tabularnewline
\hline
$Br(B^{-}\to\pi^{0}K^{-})$ & $14.7_{-4.7}^{+9.1}$ & $13.5_{-5.1}^{+3.0}$ & 11.7 & $12.5_{-3.4}^{+4.5}$ & $13.1_{-0.6-0.1}^{+0.5+0.0}$ & $12.9\pm0.5$ & $12.94_{-0.51}^{+0.52}$\tabularnewline
\hline
$A_{CP}(B^{-}\to\pi^{0}K^{-})$ & $7.8_{-2.6}^{+2.9}$ & $8.4_{-1.3}^{+1.0}$ & -3.6 & $2.2\pm2.1$ & $3.6_{-0.2-2.5}^{+0.3+2.5}$ & $3.7\pm2.1$ & $4.0\pm$2.1\tabularnewline
\hline
$Br(\bar{B}^{0}\to\pi^{+}K^{-})$ & $23.6_{-8.1}^{+15.9}$ & $21.4_{-3.0}^{+5.2}$ & 18.4 & $17.7_{-4.9}^{+6.4}$ & $19.9_{-0.6-0.0}^{+0.5+0.0}$ & $19.6\pm0.5$ & $19.57_{-0.52}^{+0.53}$\tabularnewline
\hline
$A_{CP}(\bar{B}^{0}\to\pi^{+}K^{-})$ & $4.4_{-2.0}^{+2.3}$ & $4.9_{-1.1}^{+0.8}$ & -4.1 & $-6.5\pm3.1$ & $-8.2_{-0.0-0.7}^{+0.0+0.7}$ & $-8.2\pm0.6$ & $-8.2\pm0.6$\tabularnewline
\hline
$Br(\bar{B}^{0}\to\pi^{0}\bar{K}^{0})$ & $10.2_{-3.8}^{+7.9}$ & $9.2_{-1.4}^{+2.5}$ & 8.0 & $7.4_{-2.1}^{+2.7}$ & $10.0_{-0.5-0.0}^{+0.5+0.4}$ & $9.9\pm0.5$ & $9.93\pm0.49$\tabularnewline
\hline
$A_{CP}(\bar{B}^{0}\to\pi^{0}K^{0})$ & $-3.4_{-1.8}^{+1.4}$ & $-3.7_{-0.7}^{+0.8}$ & 0.8 & $-7.9_{-1.1}^{+0.9}$ & $-1.0_{-0.1-12.3}^{+0.1+8.3}$ & $-1\pm10$ & $-1\pm10$\tabularnewline
\hline
\end{tabular}
\end{table*}

\begin{table*}
\caption{\label{tab:Direct-CP-asymmetry}The different terms of the direct CP asymmetries.
The index $i$ denotes the decay mode: $i=1$ for $B^{-}\to\pi^{-}K^{0}$
, $i=2$, for $B^{-}\to\pi^{0}K^{-}$, $i=3$ for $\bar{B}\to\pi^{-}K^{+}$,
and $i=4$ for $\bar{B}^{0}\to\pi^{0}K^{0}$. The other columns refer to the text.}
\begin{tabular}{p{1.0cm}|p{2.0cm}|p{3.0cm}|p{3.0cm}|p{3.0cm}|p{3.0cm}|p{2.0cm}}
\hline
$i$ & $N_{i}$ & 1st term $(10^{-3})$ & 2nd term $(10^{-3})$ & 3rd term $(10^{-3})$ & 4th term $(10^{-3})$ & $A_{CP}(10^{-3})$\tabularnewline
\hline
\hline
1 & $1.98$ & $-8.7$ & 0 & 0 & 0 & $-17.2$\tabularnewline
\hline
2 & $1.65$ & $-8.8$ & $31.9$ & $-1.0$ & $0.4$ & $36.9$\tabularnewline
\hline
3 & $1.94$ & $-18.0$ & $-21.9$ & $-1.0$ & $-1.0$ & $-81.3$\tabularnewline
\hline
4 & $2.25$ & $-8.3$ & $-0.8$ & $1.6$ & $2.0$ & $-12.3$\tabularnewline
\hline
\end{tabular}
\end{table*}

\begin{table*}
\caption{\label{tab:Ratios}Experimental data and the QCDF predictions for
ratios $R$, $R_{c}$, $R_{n}$.}
\begin{tabular}{p{2.0cm}|p{2.0cm}|p{2.0cm}|p{2.0cm}|p{2.0cm}|p{2.0cm}|p{2.0cm}}
\hline
Observable & 2007\citep{Baek:2007yy} & 2016\citep{Patrignani:2016xqp} & Fit & S4 & Naive & Improved\tabularnewline
\hline
\hline
$R$ & $0.89\pm0.04$ & $0.89\pm0.05$ & $0.87\pm0.04$ & 0.89 & $0.92\pm0.67$ & $0.91\pm0.26_{.}$\tabularnewline
\hline
$R_{c}$ & $1.10\pm0.07$ & $1.09\pm0.06$ & $1.10\pm0.07$ & 1.15 & $1.17\pm1.12$ & $1.19\pm0.55$\tabularnewline
\hline
$R_{n}$ & $1.00\pm0.07$ & $0.99\pm0.06$ & $1.00\pm0.06$ & 1.15 & $1.16\pm1.19$ & $1.16\pm0.42$\tabularnewline
\hline
\end{tabular}
\end{table*}

\begin{table*}
\caption{\label{tab:Amplitudes}Amplitudes calculated according to the Improved
and Fit data. Only uncertainties from the scale variables $\mu_{av}$
and $\mu_{i}$, $i=1,\cdots,4$, are included.}
\begin{tabular}{p{6.0cm}|p{3.0cm}|p{3.0cm}}
\hline
decay modes & Improved & Fit\tabularnewline
\hline
\hline
$A(B^{-}\to\pi^{-}K^{0})$(eV) & $(49.4\pm6.0)e^{-i173^{\circ}}$ & $(50.0\pm1.1)e^{-i173^{\circ}}$\tabularnewline
\hline
$A(B^{-}\to\pi^{0}K^{-})$(eV) & $(39.5\pm3.8)e^{i18^{\circ}}$ & $(38.0\pm0.9)e^{i19^{\circ}}$\tabularnewline
\hline
$A(\bar{B}^{0}\to\pi^{-}K^{+})$(eV) & $(50.8\pm5.5)e^{i15^{\circ}}$ & $(45.8\pm0.7)e^{i17^{\circ}}$\tabularnewline
\hline
$A(\bar{B}^{0}\to\pi^{0}K^{0})$(eV) & $(31.9\pm4.2)e^{-i177^{\circ}}$ & $(33.7\pm1.0)e^{-i179^{\circ}}$\tabularnewline
\hline
$A(B^{-}\to\pi^{-}K^{0})+\sqrt{2}A(B^{-}\to\pi^{0}K^{-})$ & $(12.0\pm8.0)e^{i70^{\circ}}$ & $(11.5\pm1.7)e^{i84^{\circ}}$\tabularnewline
\hline
$A(\bar{B}^{0}\to\pi^{-}K^{+})+\sqrt{2}A(\bar{B}^{0}\to\pi^{0}K^{0})$ & $(11.5\pm8.0)e^{i70^{\circ}}$ & $(13.1\pm1.6)e^{i107^{\circ}}$\tabularnewline
\hline
$r$ & $(1.04\pm1.0)e^{i0.4^{\circ}}$ & $(0.87\pm0.15)e^{-i23^{\circ}}$\tabularnewline
\hline
\end{tabular}
\end{table*}
\end{document}